\documentclass{article}

\usepackage{arxiv}

\usepackage[utf8]{inputenc} 
\usepackage[T1]{fontenc}    
\usepackage{hyperref}       
\usepackage{url}            
\usepackage{booktabs}       
\usepackage{amsfonts}       
\usepackage{nicefrac}       
\usepackage{microtype}      
\usepackage{lipsum}		
\usepackage{graphicx}
\usepackage{natbib}
\usepackage{doi}
\usepackage{color}
\usepackage{amsmath}
\usepackage{mathtools}

\makeatletter
\@addtoreset{subfigure}{figure}
\makeatother
\usepackage{xr-hyper}
\usepackage{url,hyperref,lineno,microtype,subcaption,float,graphicx,fontenc}
\usepackage[onehalfspacing]{setspace}

\usepackage{mathtools}
\usepackage{standalone}
\usepackage{tikz}
\usepackage{tkz-graph}
\usetikzlibrary{shapes,positioning}
\usetikzlibrary{arrows.meta}
\usetikzlibrary{matrix}
\usetikzlibrary{calc}

\newcommand\indep{\protect\mathpalette{\protect\independenT}{\perp}}
\def\independenT#1#2{\mathrel{\rlap{$#1#2$}\mkern2mu{#1#2}}}
\usepackage{comment}
\specialcomment{comnt}{\color{gray}}{\color{black}}
\excludecomment{comnt}

\newtheorem{defn}{Definition}
\newtheorem{expl}{Example}
\newtheorem{prop}{Proposition}

\usepackage{bibunits}

\title{Information Theoretic Measures of Causal Influences during Transient Neural Events.}

\author{ \href{https://orcid.org/0000-0002-3027-0090}{\includegraphics[scale=0.06]{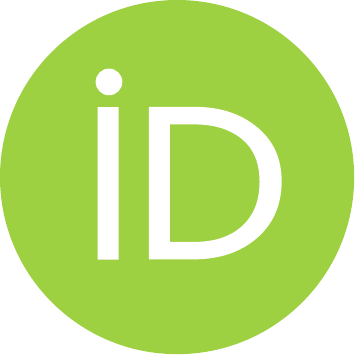}\hspace{1mm}Kaidi Shao}
	\\
	International Center for Primate Brain Research\\
	Songjiang, Shanghai, 201602, China\\
	\texttt{kaidi.shao@icpbr.ac.cn} \\
	\And
	{\hspace{1mm}Nikos K. Logothetis}\\
    International Center for Primate Brain Research\\
	Songjiang, Shanghai, 201602, China\\
	Max-Planck Institute for Biological Cybernetics\\
	Tuebingen, 72076, Germany \\
	Centre for Imaging Sciences, Biomedical Imaging Institute\\
	The University of Manchester, Manchester, UK\\
	\texttt{nikos.logothetis@icpbr.ac.cn} \\
	\And
	\href{https://orcid.org/0000-0003-0025-2323}{\includegraphics[scale=0.06]{orcid.pdf}\hspace{1mm}Michel Besserve} \\
	Department of Empirical Inference\\
	Max-Planck Institute for Intelligent Systems\\
	Tuebingen, 72076, Germany \\
	\texttt{michel.besserve@tuebingen.mpg.de} \\
}

\renewcommand{\shorttitle}{Quantifying transient causal interactions}

\begin{document}
\maketitle


\begin{abstract}
	Transient phenomena play a key role in coordinating brain activity at multiple scales, however, their underlying mechanisms remain largely unknown. A key challenge for neural data science is thus to characterize the network interactions at play during these events. 
	Using the formalism of Structural Causal Models and their graphical representation, we investigate the theoretical and empirical properties of Information Theory based causal strength measures in the context of recurring spontaneous transient events. 
	After showing the limitations of \textit{Transfer Entropy} and  \textit{Dynamic Causal Strength} in such a setting, we introduce a novel measure, \textit{relative Dynamic Causal Strength}, and provide theoretical and empirical support for its benefits. 
	These methods are applied to simulated and experimentally recorded neural time series, and provide results in agreement with our current understanding of the underlying brain circuits. 
\end{abstract}


\section{Introduction}

During both wakefulness and sleep, the mammalian brain is able to implement  numerous functions key to our survival with extraordinary reliability. This implies precise coordination of transient \textit{mechanisms} at multiple spatiotemporal scales ensuring both the synergy between brain regions contributing to the same task, and the non-interference between network activities in charge of different functions. Evidence for such transient mechanisms is provided by the variety of neural events that can be observed in brain activity across multiple structures. Such phenomena may occur in response to stimuli, as has been observed for gamma oscillations \citep{tallon1999oscillatory,fries2015rhythms}, and may play a role in the dynamic encoding of information. However, key phenomena can also occur spontaneously, as exemplified by the variety of events occurring during sleep.
These include sharp-wave ripple (SWR) complexes that occur in the hippocampus during the same sleep stages, and take the form of a slow deflection (the sharp wave, SW) superimposed with a fast short-lived oscillation (the ripple). SWR has been extensively studied and a large set of evidence supports its key role in episodic memory consolidation and the recall of previous experiences \citep{Stengel2010,Diba2007,Lee2002}. 

In order to understand how these transient phenomena operate mechanistically, causality measures based on observed neural time series can be very helpful to quantify the underlying transient influences between brain structures. Several measures of causality have been proposed, starting in the econometrics literature with Granger causality (GC) \citep{granger_1969}, relying on vector auto-regressive models. This measure can be generalized to an information-theoretic quantity: Transfer Entropy (TE) \citep{schreiber2000measuring}. In the present work, we focus on ``model-free'' quantities such as TE that are defined independently of the specific functional relationships entailed by a particular model of the dynamics. 
TE and GC have been used to assess the significance of causal links, but also the ``strength'' of these links. 
However, whether they are appropriate quantities to measure such strength is  debated \citep{stokes_2017, janzing_2013}. 

Structural Causal Models (SCM) also allow causality measures to be evaluated by their ability to reflect putative interventions targeting a specific mechanism composing the system under consideration. In this context, the relevance of causality measures has been investigated in \cite{ay2008information}, which discusses how to account for the effect of knockout experiments, and introduces a measure of \textit{information flow}, emphasizing its desirable properties. \citet{janzing_2013} provides interesting theoretical justifications for this kind of measure and extends it to define \textit{causal strength} (CS) of an arbitrary set of arrows in a graphical model. With respect to TE, information flow and CS have the benefit to be local, in the sense that it depends only on the direct causes of the observed effects and their associated mechanisms. This makes CS a good candidate to measure transient connectivity changes during non-stationary neural events, as they would be able to restrict themselves to causal influences that take place at a specific time, associated to specific arrows in the ``unrolled'' causal graph describing time-varying interactions.

However, we will argue that such a measure may not reflect well a key element for neuroscientists: the role played by transient dynamics occurring in the ``source'' region in driving events in the target region. 
Based on the potential outcome framework \citep{rubin1974estimating}, causal reasoning has also been used to provide intuitive measures of the \textit{causal impact} of a specific phenomenon happening 
at a given time point \citep{brodersen2015inferring}, by comparing it to a scenario where this phenomenon does not happen.  
This inspired us to take into account the peri-event change of signals compared to a pre-event stage as another component of causal influence. 

Therefore, in this paper, we use the lens of interventions in SCMs to propose a principled quantification of the strength of causal interactions in peri-event time series, i.e. dataset collected specifically around the times of occurrence of an identified phenomenon in neural signals. 
Based on information theoretic analyses, we assess the relevance and issues raised by a time-varying implementation of GC, TE and causal strength (DCS), and extend DCS to a novel measure, the relative DCS (rDCS), to quantify causal influences reflected by both the connectivity and the event-related change at the cause.
We show theoretically that rDCS is effective in uncovering dynamic causal influences for task-dependent events that are often accompanied with a deterministic component, as well as for spontaneous events. 
We also demonstrate how choices made for aligning peri-event time series collected across multiple occurrences of these events may  bias causality measures, and we propose a proper way to align the detected events to recover the ground truth causal direction for a uni-directionally coupled system. 
The benefits of rDCS over TE and DCS is demonstrated by both simulated toy models and neurophysiological recordings of SWRs. 
Overall, our results suggest that rDCS helps better understand the causal interactions between transient dynamical events, and thus uncover elementary mechanisms that shape brain activities. 

\section{Methods}

\subsection{General principles for the analysis of event-related causal interactions} \label{sec:principle_causality}

\subsubsection{Interventions in SCMs}\label{sec:intervene_scm}
One key question in causality is estimating the effect of manipulations of the system of interest from data, which boils down to comparing two ``worlds'' or scenarios \citep{shpitser2008complete}: the original world where no intervention is performed, and the post-intervention world. 

Both original and post-intervention worlds typically cannot be measured simultaneously (e.g. ``treatment'' and ``no treatment'' in the same patient). However, estimating their differences arguably forms the basis of causal investigations in empirical sciences, for example by performing randomized experiments on multiple instances of a system designed with mutually exclusive treatments to infer the outcome of \textit{manipulations} of this system. 
However, even performing carefully controlled experiments on close to identify instances of a system is often challenging in reality, as many physical and physiological phenomena cannot be easily reproduced or manipulated. 
This is typically the case for \textit{spontaneous} transient neural events investigated in this paper, where neurophysiological experimental techniques limit the understanding and control of their conditions of occurrence, as well as the ability to precisely modify some aspects of network activity to test assumptions on the underlying mechanisms. 

Under additional assumptions, the framework of SCMs (as briefly introduced in Supplementary Section A), can be leveraged to infer the outcome of manipulations based on observational data only. Assuming those assumptions are met (which is out of the scope of the present work), the SCM inferred from data can be modified using a family of operations named \textit{interventions} to model the pseudo-manipulation of the system described by the SCM \citep{pearl2000causality, causality_book}. 
Intervening typically refers to modifying the structural equation associated to one node in the SCM, to study the modifications it brings about in the system. 
When interventions are performed, the only affected mechanistic relations (represented by arrows in an SCM) are the ones between the intervened nodes and their parent nodes. For instance, one can impose a fixed deterministic value to a node, or that this node's variable is drawn from a given distribution, independently from other variables in the SCM  \citep[Chapter 3]{janzing_2013,correa2020calculus,causality_book}. Both such interventions lead to an intervened causal graph where the arrows between the node intervened upon and its parents are removed.  

Importantly, while an intervention modifies the graph associated to an SCM, the variables' joint distribution can still be obtained by exploiting the intervention knowledge, observational data and prior assumptions related to the unaffected conditionals. 
Mathematically, for an SCM where a directed acyclic graph ($\mathcal{G}$) is described by the following structural equations 
\[
V_j \coloneqq f_j(\textbf{PA}_j,N_j), j=1,\dots,d.
\]
$\textbf{PA}_j$ are the variables indexed by the set of parents of vertex $j$ in $\mathcal{G}$. 
Intervening on $V_k$ consists in replacing its structural assignment by a new one:
\[
V_k\coloneqq \widetilde{f_k}(\widetilde{\textbf{PA}_k},\widetilde{N_k})\,.
\] 
The resulting modified distribution $\widetilde{P}_V=P_V^{\text{ do}(V_k\coloneqq\widetilde{f_k}(\widetilde{\textbf{PA}_k},\widetilde{N_k}))}$ 
is called \textit{intervention distribution} (see e.g. \citet[chapter 6]{causality_book}). 
Meanwhile, other structural equations and the distribution of their associated exogenous variables are kept unchanged.

As an example, Figure~\ref{fig:counterfactual}A shows two uni-directionally coupled brain regions where transient events are observed and the corresponding SCM. 
To obtain an intervened system mimicking the experimental ablation of anatomical connectivity, Figure~\ref{fig:counterfactual}B shows the intervention performed in the SCM: cut the causal arrow from $X^2_{t-1}$ to $X^1_t$ and feed $X^1_t$ with an independent copy of $X^2_{t-1}$ (denoted as $X^{2'}_{t-1}$). The rationale behind this operation is that we want to suppress the dependency between the two nodes while maintaining the same level of input activity in the target node. In a context where nodes correspond to single neurons, this can be thought of as a proxy for the experiment of cutting the axon of afferent neurons, while injecting a current to maintain the baseline level of excitation in the target neuron, such that it is kept in naturalistic conditions. 

\subsubsection{
Both activity in the source region and connectivity causally influence the target region 
} \label{sec:indepen_cause_mechan}

At first glance, the aforementioned scenario seems to straightforwardly contrast the causal effect we wish to measure with a reasonable baseline. However the operation of feeding the effect node $X^1_t$ with an independent copy of the cause node $X^2_{t-1}$ at the same time $t-1$ still implicitly incorporates the influence of the event-related transient changes undergone by $X^2$ at the time $t-1$ on $X^1_t$, as the distribution of $X^2_{t-1}$ may strongly differ from what it is during baseline activity (before the event onset). 
By removing the causal link in the intervened SCM, we are measuring the influence of connectivity on the target region at the time the event happens in the ``cause region'', but do not contrast this influence to a situation where the event would not have happened. We thus argue that a better reference scenario for testing the influence of an event in a source region on a target region would both ``remove the connectivity between two regions'' and also ``remove the event-related changes in the cause region'' (Figure~\ref{fig:counterfactual}C(upper)). This would account for both the cases of stimulus-triggered events and spontaneous events, as addressed in Section~\ref{sec:insensit_rDCS} and Section~\ref{sec:align}. We refer to the \textit{causal impact} and regression discontinuity methodologies to justify how to implement this in the next section. 

\subsubsection{Causal impact and regression discontinuity to remove the existence of cause events} \label{sec:causal_impact}

\begin{figure}
	\includegraphics[width=\textwidth]{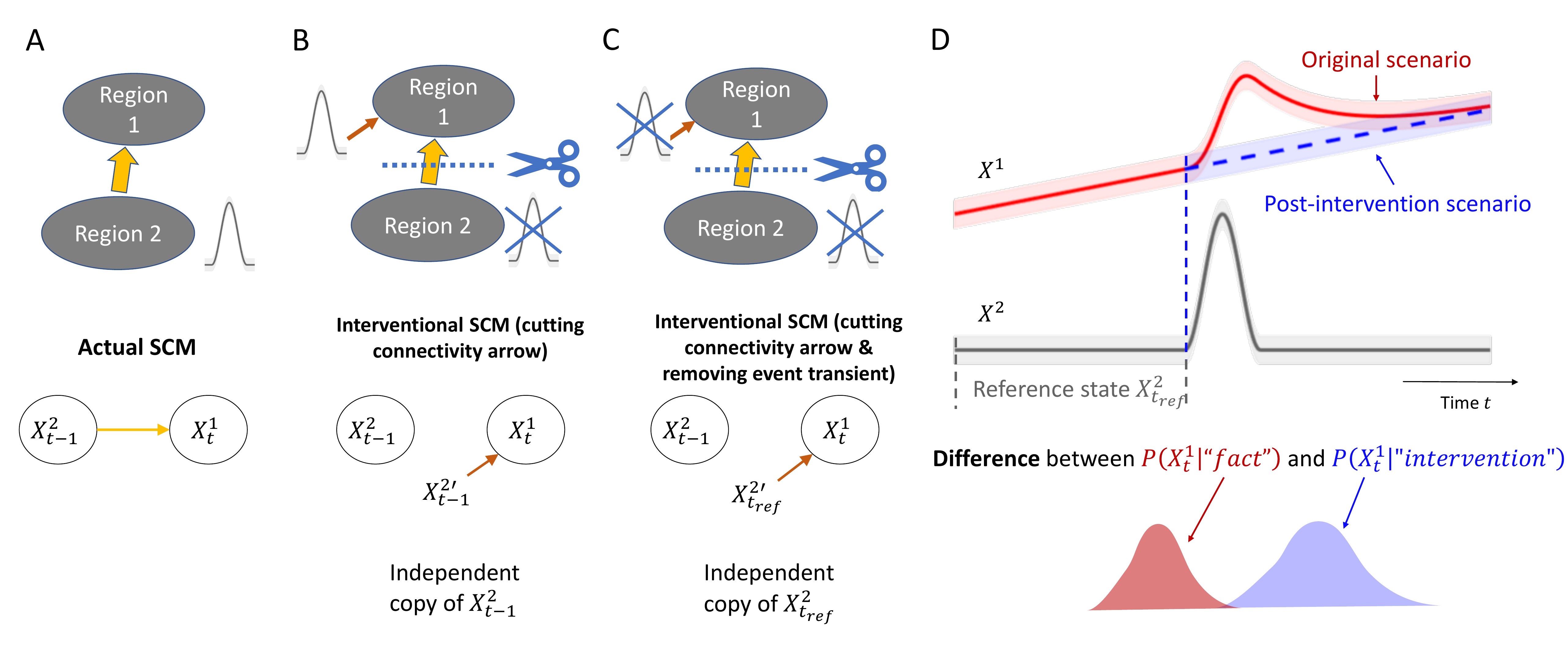}
	\caption{Analysis of event-based causality via interventions in SCMs. 
	(A) (upper) A diagram representing two brain regions with uni-directional connectivity from Region 2 to Region 1. Region 2, as the ``cause region'', exhibits transient events (grey) that influence Region 1 repetitively.
		(lower) An SCM underlying the diagram, where $X^1_t$ and $X^2_t$ denotes states of Region 1 and Region 2.
	(B) (upper) An experimental manipulation of the two-region diagram in (A): cutting the anatomical connectivity.
		(lower) A corresponding intervention of the SCM in (A) represents cutting the causal arrow and feeding the effect node $X^1_t$ with an independent copy of the cause node $X^2_t$.  
	(C) (upper) Another experimental manipulation of the two-region diagram in (A): cutting the anatomical connectivity and removing the event-based signal changes at Region 2.
		(lower) The corresponding intervention of the SCM in (A) represents cutting the causal arrow and feeding the effect node $X^1_t$ with an independent copy of a reference state of the cause node $X^2_t$. 
	(D) (upper) A time course of observed peri-event signals of Region 1 ($X^1_t$, red) and Region 2 ($X^2_t$, grey) reflecting the actual condition. 
	The blue dashed time course represents the post-intervention scenario where $X^1_t$ evolves without the influence from $X^2_t$. 
	The interval marked by grey dashed lines refers to the reference state before the occurrence of events in $X^2_t$. 
	(lower) A proper causality measure should quantify the difference between the original and post-intervention scenarios at each time point.
	} \label{fig:counterfactual}
\end{figure}

Using the potential outcome framework \citep{rubin1974estimating}, \citet{brodersen2015inferring} introduced a Bayesian approach to quantify the causal impact of an event at a given time point $n$ on an observed time series $\{y_k\}$. It relies on observed pre-event data $\{{y_{1}}, ..., {y_{n}}\}$, covariates and priors on time series parameters to extrapolate a distribution of potential outcome sample paths $\{\tilde{y_{n+1}}, ..., \tilde{y_{m}}\}$ under the counterfactual scenario that no 
event occurred. 
Comparing the posterior predictive density of these unobserved counterfactual responses to the observed time course ${y_{n+1}}, ..., {y_{m}}$ (under intervention) thus allows quantifying the effect of the event. Such contrasting strategy is also present in a variant of regression discontinuity designs in economics and social sciences. In particular, \textit{regression discontinuity in time} assesses causal effects by comparing outcomes' distributions on time intervals before and after the onset of a policy change \citep{hausman2018}. 

Both approaches contrast the properties of the models over different time intervals, one before the event and one after. 
Inspired by these works, we suggest estimating the post-interventional scenario based on the dynamics of the intervened distributions and observed distributions at reference time points where the intervention had not yet occurred (Figure~\ref{fig:counterfactual}D(upper)).

Specifically, we propose that a reference scenario where both connectivity and the influence of transient events on the cause are removed can be approximated by replacing again the variable fed into the target node, as described above, but this time it should be replaced by an independent copy of the activity in the cause region at a reference time point where no event has occurred yet (Figure~\ref{fig:counterfactual}C,D).  
We will elaborate on implementation aspects in Section~\ref{sec:rDCS}.

\subsection{Candidate time-varying causality measures} \label{sec:review_causality_measures}

\begin{figure}
	\centering
	\includegraphics[width=.9\textwidth]{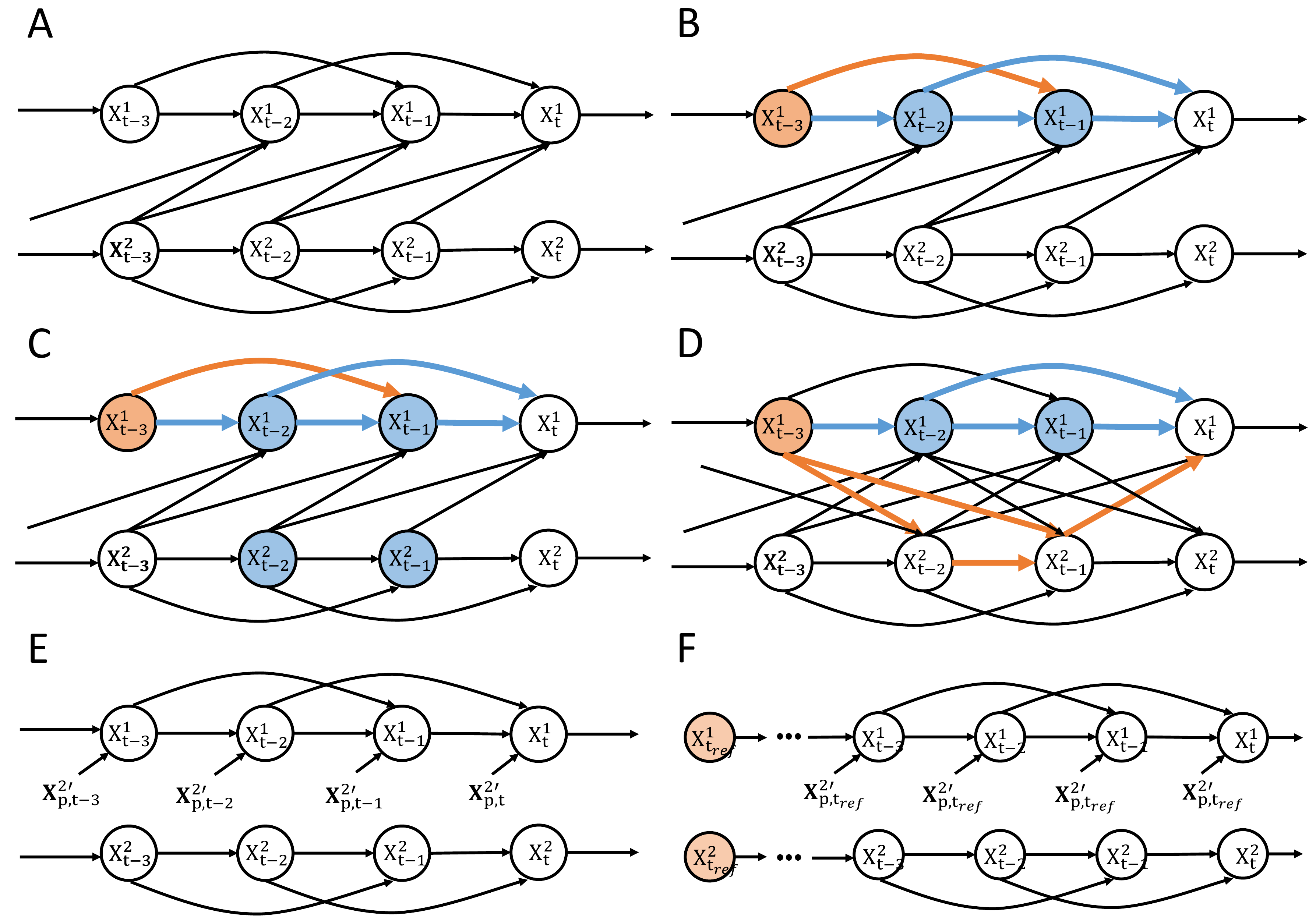}
	\caption{D-separation of bi-variate VAR(2) model.
		(A) Structural causal model of a bi-variate VAR(2) model defined in Eq.~\ref{eq:full_model_x1_inhomo} with uni-directional coupling from $X^2$ to $X^1$. 
		(B) Conditioning on both past states of $X^1$ and $X^2$ blocks all paths from $X^1_{t-3}$ to $X^1_t$. Blue nodes represents conditioned nodes while blue arrows marks blocked paths. Orange arrows marks the unblocked paths. 
		(C) Conditioning on past states of $X^1$ alone blocks all paths from $X^1_{t-3}$ to $X^1_t$ in the uni-directional case. Color codes are the same as (B). 
		(D) Conditioning on past states of $X^1$ alone does not block all paths from $X^1_{t-3}$ to $X^1_t$ in the bi-directional case. Color codes are the same as (B). 
		(E) The intervention implemented in devising CS is to break the causal arrows and send an independent copy $\boldsymbol{X}^2_{p,t}$ to $X^1_t$ at each time point.
		This diagram applies to both CS and DCS (Section~\ref{sec:DCS}).
		(F) The intervention implemented in devising rDCS is to break the causal arrows and send an independent copy of the stationary state $\boldsymbol{X}^2_{p, t_\mathit{ref}}$ to ${X}^1_t$ at each time point. } \label{fig:scm}
\end{figure}

We now present the time-varying versions of commonly-adopted causality measures and discuss their properties in the context of transient event-based causality analysis, in light of the above principles. 
The candidate measures include time-varying extensions of Granger causality (GC), Transfer Entropy (TE) and Causal Strength (CS) \citep{janzing_2013}, to address the non-stationarity of transient events. To make the comparison quantitative, time series are modeled as \textit{linear vector auto-regressive} (VAR) model, that we specify with \textit{time-inhomogeneous} (or \textit{time-varying}) coefficients to match our context. 
In the SCM framework, a bi-variate time-varying VAR model can be represented by the causal graph of Figure~\ref{fig:scm}A. 

\subsubsection{Granger causality} \label{sec:GC}

Granger causality (GC), as well as its information-theoretic extension, Transfer Entropy (TE) 
is based on Wiener's principle of causality, 
based on which \citet{granger_1969} defines (Granger-)causality from $X^2$ to $X^1$ if knowledge of  $\boldsymbol{X}^2_{p,t}$, in addition to $\boldsymbol{X}^1_{p,t}$, will allow better prediction of $X^1_t$. 
This can be interpreted as a comparison between two prediction scenarios:
\vspace*{-.3cm}
\begin{itemize}
	\item Scenario 1: predict $X^1_t$ with both $\boldsymbol{X}^1_{p,t}$ and $\boldsymbol{X}^2_{p,t}\,$,
	\item Scenario 2: predict $X^1_t$ with only $\boldsymbol{X}^1_{p,t}\,$,
\end{itemize}
\vspace*{-.3cm}
where $\boldsymbol{X}^1_{p,t}$ and $\boldsymbol{X}^2_{p,t}$ refer to the respective $p$  previous past points of each time series, without further specification, such that in our notation $p$ can be potentially infinite. 

The VAR model describing the first scenario is referred to as the \textit{full model} \citep{geweke_1984}, where the first variable $X^1$ is dependent on both variables $X^1$ and $X^2$:
\begin{equation} \label{eq:full_model_x1_inhomo}
	X^1_t = \mathbf{a}^{\top}_t \boldsymbol{X}^1_{p,t} + \mathbf{b}^{\top}_t \boldsymbol{X}^2_{p,t} +\eta^1_t\,,\quad\eta^1_t \sim \mathcal{N}(k^1_t,\,\sigma_{1,t}^2)\,.
\end{equation}
where the $t$ subscript in all parameters $(\mathbf{a}_t,\,\mathbf{b}_t,\,k^1_t,\,\sigma_{1,t}^2)$ comes from our time-inhomogeneity assumptions and is not standard in the GC literature. An estimate of the innovation variance of ${X^1_t}$ ($\sigma_{1,t}^2$ in Eq.~\ref{eq:full_model_x1_inhomo}) is the mean squared residual error ($\hat{\sigma}_{1,t}$) of the forecast of $X^2_t$ under the assumption that both $\boldsymbol{X}^1_{p,t}$ and $\boldsymbol{X}^2_{p,t}$ contribute to $X^1_t$. 
Under Scenario 2 where $X^1_t$ is predicted only by $\boldsymbol{X}^1_{p,t}$, we have a \textit{reduced model}
\begin{equation} \label{eq:reduced_model_Y}
X^1_t = \mathbf{a}'^{\top} \boldsymbol{X}^1_{p',t} +{\eta^{1}}'\,,\quad {\eta^{1}}' \sim \mathcal{N}(k^1,\,{\sigma'}_{1,t}^2)\,.
\end{equation}
where the model order $p'$, the coefficient $\mathbf{a}'$, 
the innovations mean $k^1$ and innovations variance ${\sigma'}_{1}^2$ are different from the corresponding terms in Eq.~\ref{eq:full_model_x1_inhomo} and should be re-estimated. 

If $X^2$ Granger-causes $X^1$, then the full model should fit the data more accurately compared to the reduced model as measured by the estimated variance $\widehat{\sigma'}_{1,t}^2$, which should be larger than the one of $\widehat{\sigma}_{1,t}^2$. 
Then the Granger causality can be defined as the log ratio of the residual variance between the reduced model and the full model, which leads to estimating the magnitude of Granger causality as
\begin{equation} \label{eq:GC_definition_homo}
\text{GC}(X_t^2 \rightarrow X_t^1)= \frac{1}{2}\log\left( \frac{{\widehat{\sigma'}_{1,t}}^2}{{\widehat{\sigma}_{1,t}}^2}\right)\,,
\end{equation}
where the factor $1/2$ is chosen for consistency with TE (see Section~\ref{sec:TE}). 
While the above linear VAR model is the most widely used, Granger causality has been extended to non-linear models following the same predictive approach (e.g. \cite{marinazzo2008kernel,marinazzo2011nonlinear,diks_2016}).

\subsubsection{Transfer Entropy} \label{sec:TE}
TE is an information-theoretic implementation of Wiener's principle, where the performance of prediction between the above two scenarios is quantified with conditional entropy. 
In information theory, the conditional entropy $H(X|Y)=\mathbb{E}_y[H(X|Y=y)]$, measures the amount of information needed to describe the outcome of random variable $X$ given that the value of another random variable $Y$. In the context of Wiener's principle, this can be used as a generalized way of quantifying the quality of the prediction of future values based on past ones: the larger $H(\mbox{future}|\mbox{past})$, the worse the quality of the prediction is.

Transfer Entropy (TE) quantifies to which amount $X^2$ is Granger causes $X^1$ and is defined as
\begin{eqnarray}\label{eq:TE_entrop}
	\text{TE}(X_t^2 \rightarrow X_t^1) & =  & H(X^1_t|\boldsymbol{X}^1_{p,t})-H(X^1_t|\boldsymbol{X}^1_{p,t},\boldsymbol{X}^2_{p,t})\,.
\end{eqnarray}
Interestingly, using the Kullback-Leibler (KL) divergence $D_{KL}$ between two probability densities
$D_{KL}(p||q) = \int p(x) \log \frac{p(x)}{q(x)} dx\,,
$
TE can be rewritten as an expected  KL-divergence between the corresponding conditional probabilities, thereby contrasting the two above mentioned scenarios:
\begin{equation} \label{eq:TE_KL}
\text{TE}(X_t^2 \rightarrow X_t^1) = \mathbb{E}_{(\boldsymbol{X}^1_{p,t},\boldsymbol{X}^2_{p,t})}\left[D_{KL}\left(p(X^1_t|\boldsymbol{X}^1_{p,t}, \boldsymbol{X}^2_{p,t})\|p(X^1_t|\boldsymbol{X}^1_{p,t})\right)\right]\,.
\end{equation}
As noticed in \cite{barnett2009granger}, under stationary Gaussian VAR assumptions the analytic expression of  Gaussian entropy 
applied to Eq.~\ref{eq:TE_entrop} leads to
$	\text{GC}(X_t^2 \rightarrow X_t^1)= \text{TE}(X_t^2 \rightarrow X_t^1)
$
in the limit of unbiased variance estimation, such that TE appears as a strict generalization of GC, and can be estimated by GC in the context of Gaussian VAR models.
TE and GC statistics are two commonly used measures of causal strength for investigating interactions between brain regions (e.g., \citet{wibral_2013,besserve_2010,besserve_2015}). 
Based on the observational conditional distribution of the neural signals being analyzed, these two measures estimate a quantity that is easily interpretable from a forecasting perspective.
However, they have some limitations with regard to their interpretability as interventions in the SCM framework and in the time varying setting that interests us in this paper.

A key issue is that the reduced model ignores but \textit{does not remove} the influence of past values of $\boldsymbol{X}^2$ ($\boldsymbol{X}^2_{p,t}$) on $\boldsymbol{X}^1_t$ by marginalizing with respect to them. 
It can be shown that such change does not preserve the SCM structure, and leads to violations of the Markov properties due to the implicit dependency on the mechanisms relating 
$\boldsymbol{X}^2_{p,t}$ and $\boldsymbol{X}^1_{p,t}$, which manifest themselves through the $p(\boldsymbol{X}^2_{p,t}|\boldsymbol{X}^1_{p,t})$ term of the marginalization equation \citep{ay2008information, janzing_2013}:
\begin{equation}\label{eq:TEmargin}
p(X^1_t|\boldsymbol{X}^1_{p,t})
= \int p(X^1_t|\boldsymbol{X}^1_{p,t},\boldsymbol{X}^2_{p,t})p(\boldsymbol{X}^2_{p,t}|\boldsymbol{X}^1_{p,t})
\mathrm{d}\boldsymbol{X}^2_{p,t}\,.
\end{equation}
As a consequence, the reduced model cannot be interpreted as an intervention on the original SCM that would result in a model where arrows associated to the causal influence of interest would be removed. 

Besides, TE estimation is non-local. 
While one can exploit classical model order selection techniques (e.g. Akaike Information Criterion \citep{akaike_1974, akaike1998information} and Bayesian Information Criterion (BIC,  \cite{gideon1978estimating, KaidiBIC}) to select the best order for the full model, in case of bi-directional coupling, the reduced model of Eq.~\ref{eq:reduced_model_Y} is misspecified (in a generic case) for any finite order.
This can be easily seen by exploiting the $d$-separation criterion (see Supplementary Section A), as illustrated in Figure~\ref{fig:scm}.
Figure~\ref{fig:scm}B shows the estimation in the full model, where conditioning on both $\boldsymbol{X}^1_{p,t}$ and $\boldsymbol{X}^2_{p,t}$ blocks all the paths from $X^1_{t-3}$ to $X^1_t$ such that 
$X^1_{t-3}$ and $X^1_t$ are conditionally independent.
For such a uni-directionally-coupled system, a finite order for the reduced model also guarantees such conditional independence, as seen in Figure~\ref{fig:scm}C where all paths are blocked by conditioning.
However, in the same system with bi-directional coupling, for any $k>p$ (i.e. $k>2$), there is always a path from $X^1_{t-k}$ to $X^1_{t}$ going through nodes of $X^2$ that is unblocked by $(X^1_{t-p},\,\cdots, X^1_{t-1})$. 
As Figure~\ref{fig:scm}D shows, 2 paths from $X^1_{t-3}$ to $X^1_t$ are not blocked by conditioning on $\boldsymbol{X}^1_{p,t}$. 
Under faithfulness assumptions, this implies that there is conditional dependence between $X^1_t$ and its remote past samples, no matter how many finite past states we are conditioning on. 
This further implies that to minimize the forecast error of $X^1_t$ in the reduced model one should ideally exploit the past information of this time series up to $p=+\infty$.

This issue has been both raised and addressed in the literature, in particular by resorting to Autoregressive Moving Average models and state space models for defining an appropriate reduced model (e.g. \citep{barnett_2015, solo_2016}). 
However, this remains an important limitation when extending TE to time-varying versions, where the model is assumed to be stationary at best locally in time. 
For example, when defining a non-stationary VAR model as Eq.~\ref{eq:full_model_x1_inhomo}, we assume a different linear model in each 1-point time window.
The non-locality of TE is particularly problematic for such a time-varying model assumption because of the implicit influence of past activities on this quantity. 

\subsubsection{Dynamic causal strength}\label{sec:DCS}
\label{sec:motivation}
To overcome the limitation of TE and GC, \cite{ay2008information} has proposed a measure of \textit{information flow} to quantify the influence of some variables on others in a system, which has been further studied and generalized in \cite{janzing_2013} as a measure of the \textit{Causal Strength} (CS) of an arbitrary set of arrows in a graphical model. 
In the present paper, we define CS in the context of time-inhomogeneous vector autoregressive processes and their associated unrolled causal graph, and thus call it \textit{Dynamic Causal Strength} (DCS).

DCS can be naturally defined using the SCM interventional formalism  (\cite{pearl2000causality, causality_book}, see also Supplementary and Section~\ref{sec:intervene_scm}). 
Briefly, interventions are performed on nodes in order to remove the specific arrows from the causal graph whose influence we wish to quantify.
In agreement with \cite{ay2008information} and \cite{janzing_2013}, in the context of inhomogeneous VAR models (as illustrated in Figure.~\ref{fig:scm}A), an appropriate intervention to address the causal inference from $\boldsymbol{X}^2_t$ to $X^1_t$ can be designed as the following soft intervention (shown in Figure.~\ref{fig:scm}E): 
\textit{replace the arrow $\boldsymbol{X}^2_{p,t}\rightarrow X^1_t$ by an arrow injecting instead ${\boldsymbol{X}^2_{p,t}}'$, an independent copy of  $\boldsymbol{X}^2_{p,t}$ with the same marginal}.
Importantly, compared to \cite{janzing_2013} but in line with \cite{ay2008information}, we propose to replace the multivariate vector $\boldsymbol{X}^2_{p,t}$ with a copy without enforcing independence between the components of this vector, in order to preserve the dependency between the successive past time points of $X^2$, as those are often strongly correlated in practice. 
The intervention distribution $p^{DCS}$ models the post-interventional world after removing the causal arrow from $\boldsymbol{X}^2_{p,t}$ to $X^1_t$ and results in the entailed conditional probability  

\begin{eqnarray*}
p^{{DCS}}
(X^1_t|\boldsymbol{X}^1_{p,t})
=	p^{do(X^1_t \coloneqq f(\boldsymbol{X}^1_{p,t},{\boldsymbol{X}^2_{p,t}}^\prime, \eta^1_t))}
	(X^1_t|\boldsymbol{X}^1_{p,t}, \boldsymbol{X}^2_{p,t})
	 =\!\! \int p(X^1_t|\boldsymbol{X}^1_{p,t},\boldsymbol{X}^2_{p,t})p({\boldsymbol{X}^2_{p,t}})
	\mathrm{d}{\boldsymbol{X}^2_{p,t}}\,,
\end{eqnarray*}
which does not depend on $p(\boldsymbol{X}^2_{p,t}|\boldsymbol{X}^1_{p,t})$ anymore, in comparison to Eq.~\ref{eq:TEmargin}. DCS then quantifies the KL divergence between the distributions of $X^1_t|(\boldsymbol{X}^1_{p,t}, \boldsymbol{X}^2_{p,t})$ obtained in both worlds, such that
\begin{equation}\label{eq:DCS}
	\text{DCS}(X_t^2 \rightarrow X_t^1) 
	=  \mathbb{E}_{\boldsymbol{X}^1_{p,t},\boldsymbol{X}^2_{p,t}}\left[D_{KL}(p(X^1_t|\boldsymbol{X}^1_{p,t}, \boldsymbol{X}^2_{p,t}) \mid
	p^{DCS}(X^1_t|\boldsymbol{X}^1_{p,t}))\right].
\end{equation}

A parametric formulation under linear Gaussian model assumptions is given in Supplementary Section D.5.

\subsection{Near deterministic behavior of TE and DCS} \label{sec:dcs_issues}
The analysis of transient neural events leads us to analyze signals that have limited stochasticity in several respects: on the one hand, strongly synchronized oscillatory signals can be represented by VAR models with low innovation variance, relative to the variance of the measured signal. Moreover, when a study focuses on a reproducible type of transient pattern, it often has a reproducible component, with little variability across collected trials. Such a situation can be modeled with a time-varying deterministic innovation, exhibiting strong variation of its mean across time, but no or little variance. We investigate the theoretical properties of TE and DCS in this regime, showing a benefit of DCS with respect to TE, but also remaining limitations.  

\subsubsection{TE behavior for strongly synchronized signals}
Besides, it has also been pointed out that the definition of TE in Eq.~\ref{eq:TE_KL} has some other non-intuitive implications \citep{ay2008information,janzing_2013}. 
In particular, there are situations in which $\text{TE}(X^2\rightarrow X^1)$ almost vanishes, although the influence is intuitively clear. 
How frequent are the practical situations in which we have these detrimental effects is unclear; however, theoretical analysis suggests that this can happen when the time series are strongly correlated. 

To see this, we can derive with Eq.~\ref{eq:TE_KL} in the case where $X^2$ is a deterministic function of $X^1$ such that TE vanishes.
Take the special case where $X^2_t$ is proportional to $X^1_t$ such that $X^2_t = k X^1_t$, representing a time-wise synchronization of the two signals, the conditional variance will be
\[
\Sigma_{\boldsymbol{X}^2_p |\boldsymbol{X}^1_p} = \Sigma_{\boldsymbol{X}^2_p} - \Sigma_{\boldsymbol{X}^2_p \boldsymbol{X}^1_p}\Sigma_{\boldsymbol{X}^1_p}^{-1}\Sigma_{\boldsymbol{X}^1_p \boldsymbol{X}^2_p} = \Sigma_{\boldsymbol{X}^2_p} - k\Sigma_{\boldsymbol{X}^2_p} \cdot (\frac{1}{k^2}\Sigma_{\boldsymbol{X}^2_p}^{-1}) \cdot k\Sigma_{\boldsymbol{X}^2_p} = 0
\]    
Plugging into Eq.S8 in Supplementary Section D.4 yields, 
\[
\text{TE}(X_t^2 \rightarrow X_t^1)  = \log \frac{\mathbf{b}_t^\top \text{Cov}[\boldsymbol{X}^2_{p,t}|\boldsymbol{X}^1_{p,t}]\mathbf{b}_t +{\sigma_{1, t}}^2}{{\sigma_{1, t}}^2} = \log{\frac{{\sigma_{1,t}}^2}{{\sigma_{1,t}}^2}} = \log1 = 0
\]
However, a strong correlation between two observed time series does not necessarily imply that causal interactions between them are weak, from an SCM perspective.
We will investigate this case in Section~\ref{sec:synchron_signals} and compare with the results of DCS to show that DCS does not suffer from this non-intuitive vanishing problem. 
\subsubsection{ Insensitivity of TE and DCS to deterministic perturbations} \label{sec:te_insensit}

While several intuitive properties make DCS a good candidate to quantify causal influences, we exhibit a counterintuitive property common to TE and DCS in the context of peri-event time series.
Transient neural events are mainly investigated in two types of analyses: 1) stimulus-triggered (or response-triggered) data that are temporally aligned by task (or response) onset and 2) event-triggered data where occurrences of a type of brain-activity pattern are detected along the time course of the recordings (manually or algorithmically) and used to create peri-event trials. 

In both cases, neural activities are likely to have a deterministic component appearing in the peri-event ensembles, due the similarity of the response to successive stimuli in case 1), or due to the similarity of the neural patterns detected in the recordings in case 2).  
Here we will show that, in a linear setting, TE and DCS are insensitive to such a deterministic component.
Specifically, TE and DCS values are unaffected by interventions on the innovations' mean at any time point.

First, we define what is referred to as deterministic perturbation. Consider an example bi-variate VAR(1) model in the following form
\begin{subequations}\label{eq:detinterexple}
	\begin{alignat}{3}
		\label{eq:preferred_VAR_X}
		&X^1_t && \quad\coloneqq &&\quad a X^1_{t-1} + b X^2_{t-1}  +\eta^1_t,\\
		\label{eq:preferred_VAR_Y}
		&X^2_t && \quad\coloneqq && \quad  \eta^2_t,
	\end{alignat}
\end{subequations}
with $a,b\neq 0$ and a stationary innovation for $X^1$, $\eta^1_t\sim\mathcal{N}(0,1)$, but a non-stationary innovation for $X^2$, $\eta^2_t\sim\mathcal{N}(\alpha\delta_{t,\, t_0},1)$, with \[
\delta_{t, t_0}=
\begin{cases}
	1,& \mbox{for } t=t_0\,,\\
	0,& \mbox{otherwise}\,. 
\end{cases}
\]
When varying $\alpha$, this models a (soft) intervention on the second time series.  
Then it can be easily shown that the expected time course of $X^1$ is
\[\mathbb{E} \left[X^1_t\right] = \begin{cases}
	\alpha b a^{t-t_0+1},&t\geq t_0+1\\
	0,& \mbox{otherwise.}
\end{cases}\]
This witnesses the causal influence of $X^2_{t_0}$ on values of $X^1_t$ at subsequent times, which for large $\alpha$ results in large deviations from the baseline expectation of $X^1_t$ for $t$ prior to $t_0$. Intuitively, one may expect that a quantification of the magnitude (strength) of the causal influence of $X^2$ on $X^1$ should be larger for larger $\alpha$, as a transient of larger magnitude propagates from $X^2$ to $X^1$. From a neuroscientific perspective, this could model an experimental setting where one brain region is electrically stimulated with increasing strength to detect whether it is anatomically connected to another. Obviously, the magnitude of the stimulation is expected to be critical to elicit a response in the target region. However, TE and DCS actually turn out to be insensitive to such stimulation. 

We will show this in the more general setting of the VAR($p$) model of Eq.~(\ref{eq:full_model_x1_inhomo}) and Supplementary Eq.S7. Consider the intervention at time $t_0$ that transforms $\eta_{t_0}$ to $\eta_{t_0}+\alpha$. To compute the intervention distribution of the new variables denoted $(\tilde{X}^1,\tilde{X}^2)$ changes with respect to the distribution of the original variables, we can examine the difference with respect to $(X^1, X^2)$ that has the same innovations, except for $\eta^1_{t_0}$ for which we remove a constant $\alpha$. $(X^1, X^2)$ is then distributed according to the original distribution (before intervention), and the difference $(U, V)=(\tilde{X}^1-X^1,\tilde{X}^2-X^2)$ follows the equations
\begin{alignat*}{3} 
	&U_t &&=\,&& \mathbf{a}^\top \mathbf{U}_{p,t} + \mathbf{b}^\top \mathbf{V}_{p,t} \\ 
	&V_t &&=\,&& \mathbf{c}^\top \mathbf{U}_{p,t} + \mathbf{d}^\top \mathbf{V}_{p,t} +\delta_{t\,t_0}  
\end{alignat*}
which is a deterministic difference equation with a unique solution making $\boldsymbol{X}$ and $\tilde{\boldsymbol{X}}$ coincide before the intervention\footnote{Because initial conditions of this deterministic linear system are set to zero before the intervention at $t_0$} $(U_t\,,V_t)$. As a consequence, the intervention distribution $\widetilde{P}$ is a shifted version of the original distribution:
\[
\widetilde{P}(X^1_t,\boldsymbol{X}^1_{p,t},\boldsymbol{X}^2_{p,t})
= P(X^1_t-U_t,\boldsymbol{X}^1_{p,t}-\mathbf{U}_{p,t},\boldsymbol{X}^2_{p,t}-\mathbf{V}_{p,t})
\]	
which implies the same for conditional marginal distributions, e.g.
\[
\widetilde{P}(X^1_t|\boldsymbol{X}^1_{p,t})=P(X^1_t-U_t|\boldsymbol{X}^1_{p,t}-\mathbf{U}_{p,t},\boldsymbol{X}^2_{p,t}-\mathbf{V}_{p,t})
\]
As a consequence TE on the intervention distribution writes
\begin{multline*}
	\text{TE}(\widetilde{X}^2_t\rightarrow \widetilde{X}^1_t) 
	= \int 
	\widetilde{p}(X^1_t,\boldsymbol{X}^1_{p,t},\boldsymbol{X}^2_{p,t}) \log \frac{\widetilde{p}(X^1_t|\boldsymbol{X}^1_{p,t}, \boldsymbol{X}^2_{p,t})}{ \widetilde{p}(X^1_t|\boldsymbol{X}^1_{p,t})}\, \mathrm{d} X^1_t \mathrm{d} \boldsymbol{X}^1_{p,t} \mathrm{d} \boldsymbol{X}^2_{p,t}\\
	= \int {p}(X^1_t\!-\!U_t,\boldsymbol{X}^1_{p,t}\!-\!\mathbf{U}_{p,t},\boldsymbol{X}^2_{p,t}\!-\!\mathbf{V}_{p,t}) \log\!\! \frac{{p}(X^1_t\!-\!U_t|\boldsymbol{X}^1_{p,t}\!-\!\mathbf{U}_{p,t}, \boldsymbol{X}^2_{p,t}\!-\!\mathbf{V}_{p,t})}{ {p}(X^1_t\!-\!U_t|\boldsymbol{X}^1_{p,t}\!-\!\mathbf{U}_{p,t})} \mathrm{d} X^1_t \mathrm{d} \boldsymbol{X}^1_{p,t} \mathrm{d}\boldsymbol{X}^2_{p,t}\\
	= \int{p}(X^1_t,\boldsymbol{X}^1_{p,t},\boldsymbol{X}^2_{p,t}) \log \frac{{p}(X^1_t|\boldsymbol{X}^1_{p,t}, \boldsymbol{X}^2_{p,t})}{ {p}(X^1_t|\boldsymbol{X}^1_{p,t})}\,\mathrm{d} X^1_t \mathrm{d}\boldsymbol{X}^1_{p,t} \mathrm{d}\boldsymbol{X}^2_{p,t}= \text{TE}({X^2_t}\rightarrow {X^1_t})\,.
\end{multline*}
The same reasoning can be applied to DCS leading to invariance as well (see Supplementary Section B).

As a consequence, such deterministic causal influences cannot be detected by TE or DCS for a broad class of models. 
This result is not what we would expect from a measure of influence, because in the above example of Eq.~\ref{eq:detinterexple}, setting a large $\alpha$ intuitively leads to a large influence of $X^1$ on $X^2$ provided $c\neq 0$. Provided that TE and DCS can be made arbitrarily small by reducing $\Sigma_t$, TE and DCS would detect no influence despite this strong effect on the mean of $X^2_t$. As elaborated above, this is in contrast to what would be expected in the neuroscientific context, and directly relates to the observational, event-related setting that we investigate: the deterministic component is due to the alignment of the data with respect to an event of interest, and we do not have a different condition to contrast the occurrence of this event with what would have happened in its absence. This analysis calls for building a synthetic baseline condition that would allow deterministic changes to be detected.  

\subsection{A novel measure: relative Dynamic Causal Strength} \label{sec:rDCS}

\subsubsection{Motivation}

{Therefore, following the guidelines for event-based causality (presented in Section~\ref{sec:principle_causality}), we propose a novel measure, the relative Dynamic Causal Strength (rDCS), as a modification of DCS. 
This measure aims at taking into account the influence of event-based changes in the cause signals independent from the connectivity (the mechanism), and notably those driven by deterministic exogenous inputs.}
In the specific problem we are investigating, the cause is the past states of $X^2$ as $\boldsymbol{X}^2_{p,t}$, while the mechanism can be represented by the model in Eq.~\ref{eq:full_model_x1_inhomo} and symbolized by the corresponding causal arrow in the SCMs.
In the measures we have introduced so far, DCS only exploits the case where the causal arrow is deleted as a post-intervention scenario but does not address the change in the cause itself.    

In the case where $X^2$ experiences a deterministic exogenous input in a transient window, the cause increases significantly; thus, intuitively, the causal effect should also be enhanced even if the causal arrow remains the same (i.e., the coefficient $\mathbf{b}$ stays unchanged).
Apart from intervening on the causal arrow, further intervention can be implemented on the cause node to construct a post-intervention scenario where the cause receives no time-varying innovations. 
Therefore, inspired by causal impact (Section~\ref{sec:causal_impact}) which characterizes the difference between the current state and a baseline state, we propose (additionally to DCS) to  replace the marginal of $\boldsymbol{X}^2_{p,t}$ by the marginal $\boldsymbol{X}^2_{p, t_\mathit{ref}}$ for a reference period $t_\mathit{ref}$.
The reference period $t_\mathit{ref}$ is typically chosen to be a stationary period before the occurrence of the transient deterministic perturbations and statistics of $\boldsymbol{X}^2_{p, t_\mathit{ref}}$ can be averaged by statistics of $\boldsymbol{X}^2_{p, t}$ within this period. 
This leads to the \textit{relative Dynamical Causal Strength} (rDCS)
\begin{multline}\label{eq:rDCS}
\text{rDCS}(X_t^2 \rightarrow X_t^1) = \\ 
\mathbb{E}_{(\boldsymbol{X}^1_{p,t},\boldsymbol{X}^2_{p,t})}
\left[D_{KL}(p(X^1_t|\boldsymbol{X}^2_{p,t}, \boldsymbol{X}^2_{p,t})
\|
p^{do(X^1_t \coloneqq f(\boldsymbol{X}^1_{p,t},\boldsymbol{X}^2_{p, t_\mathit{ref}}, \eta^1_t))}(X^1_t|\boldsymbol{X}^1_{p,t},\boldsymbol{X}^2_{p,t}))\right]
\end{multline}
while 
\begin{equation} \label{eq:rdcs_counterfactual}
	p^{do(X^1_t \coloneqq f(\boldsymbol{X}^1_{p,t},\boldsymbol{X}^2_{p, t_\mathit{ref}}, \eta^1_t))}(X^1_t|\boldsymbol{X}^1_{p,t},\boldsymbol{X}^2_{p,t}) = 
	\int p(X^1_t|\boldsymbol{X}^1_{p,t}, {\boldsymbol{X}^2_{p,t_{ref}}})p({\boldsymbol{X}^2_{p,t_{ref}}})
	\mathrm{d}{\boldsymbol{X}^2_{p,t_{ref}}}
\end{equation}
The implementation of rDCS given a VAR model is derived in Supplementary Section D.6.

Intuitively, the \textit{relativeness} originates from the comparison between the current past states $\boldsymbol{X}^2_{p,t}$ and the reference past states $\boldsymbol{X}^2_{p, t_\mathit{ref}}$ .
It is then natural to predict that in the uni-directional case, $\text{rDCS}(X^2\rightarrow X^1)=\text{DCS}(X^2\rightarrow X^1)$ for any reference time $t_\mathit{ref}$ if $X^2$ is stationary because stationarity implies that the marginal distributions of $\boldsymbol{X}^2_{p, t_\mathit{ref}}$ and $\boldsymbol{X}^2_{p, t}$ are identical.
As a particular case, this result implies that a transient loss of causal link from $X^2$ to $X
^1$ will lead to $\text{rDCS}=0$, while for a stationary bivariate system, $\text{DCS}=\text{rDCS}$ is constant.

\subsubsection{Sensitivity of rDCS to deterministic perturbations} \label{sec:insensit_rDCS}

{The definition of rDCS implies sensitivity to deterministic perturbations.
Indeed, taking the example in Section~\ref{sec:te_insensit}, the reference state $\boldsymbol{X}^2_{p,t_{ref}}$ is unaffected by the deterministic perturbation.  Consequently, the translational invariance does not hold for the intervention distribution because}
\begin{multline*}
	\text{rDCS}(\widetilde{X}^2_t\rightarrow \widetilde{X}^1_t) 
	= \int 
	\widetilde{p}(X^1_t,\boldsymbol{X}^1_{p,t},\boldsymbol{X}^2_{p,t}) 
	\log 
	\frac{\widetilde{p}(X^1_t|\boldsymbol{X}^1_{p,t}, \boldsymbol{X}^2_{p,t})}{ \int \widetilde{p}(X^1_t|\boldsymbol{X}^1_{p,t},{\boldsymbol{X}^2_{p,t_{ref}}})\widetilde{p}({\boldsymbol{X}^2_{p,t_{ref}}})
		\mathrm{d}{\boldsymbol{X}^2_{p,t_{ref}}}
	}\, \mathrm{d} X^1_t \mathrm{d} \boldsymbol{X}^1_{p,t} \mathrm{d} \boldsymbol{X}^2_{p,t}\\
	= \int {p}(X^1_t\!-\!U_t,\boldsymbol{X}^1_{p,t}\!-\!\mathbf{U}_{p,t},\boldsymbol{X}^2_{p,t}\!-\!\mathbf{V}_{p,t}) 
	\log\!\! 
	\frac{{p}(X^1_t\!-\!U_t|\boldsymbol{X}^1_{p,t}\!-\!\mathbf{U}_{p,t}, \boldsymbol{X}^2_{p,t}\!-\!\mathbf{V}_{p,t})}
	{\int p(X^1_t-U_t|\boldsymbol{X}^1_{p,t}-\mathbf{U}_{p,t},{\boldsymbol{X}^2_{p,t_{ref}}})p({\boldsymbol{X}^2_{p,t_{ref}}})
		\mathrm{d}{\boldsymbol{X}^2_{p,t_{ref}}}}\\
	\mathrm{d} X^1_t \mathrm{d} \boldsymbol{X}^1_{p,t} \mathrm{d}\boldsymbol{X}^2_{p,t}\\
	\neq \int{p}(X^1_t,\boldsymbol{X}^1_{p,t},\boldsymbol{X}^2_{p,t}) 
	\log 
	\frac{{p}(X^1_t|\boldsymbol{X}^1_{p,t},\boldsymbol{X}^2_{p,t})}
	{\int p(X^1_t|\boldsymbol{X}^1_{p,t},{\boldsymbol{X}^2_{p,t}})p({\boldsymbol{X}^2_{p,t}})
		\mathrm{d}{\boldsymbol{X}^2_{p,t}}}
	\, \mathrm{d} X^1_t \mathrm{d} \boldsymbol{X}^1_{p,t} \mathrm{d} \boldsymbol{X}^2_{p,t}
	= \text{rDCS}({X^2_t}\rightarrow {X^1_t})\,,
\end{multline*}
because the denominators do not allow equating the integrated terms by change of variables in the generic case. 
Therefore rDCS is capable of uncovering transient causal influences between stimulus-triggered events exhibiting a deterministic waveform.

\subsection{Alignment for spontaneous events} \label{sec:align}
The relevance of peri-event time-varying causal analysis using the proposed rDCS, as well as TE and DCS, depends on the modeling assumptions of peri-event data. In particular, we assume that the samples at a given peri-event time point are sampled \textit{i.i.d.} across trials form the same ground truth distribution \citep{KaidiBIC} associated to the neural events we want to study.  
This is easily satisfied for stimulus-evoked events, {as addressed in Section~\ref{sec:te_insensit} and Section~\ref{sec:insensit_rDCS}}, with a intrinsic reference time for occurrence (i.e., the triggering time). 
However, analyzing spontaneous events, like the transient events during sleep, requires a detection procedure to locate their occurrence. 
This commonly involves the procedures of transforming the original signal into a detection signal that amplifies the event-related features (i.e. by filtering or template matching) and finding the events where the detection signal is over a certain threshold.
Specifically, the events detected in this way are often aligned by the local peaks of peri-event signals, and this alignment may not reflect perfectly the ground truth distribution of the events. 

While alignment may seem a trivial step at first blush, its influence on VAR model estimation turns out to be critical. This could lead to biased estimation of event statistics and peri-event dynamics (due to selecting only data over threshold and gathering local maxima together), resulting in misleading characterization of causal interactions (e.g. wrong causal directions). 
Thus we will discuss here how to align the events appropriately such that the causal direction between transient events can be better identified with the proposed rDCS. 

\begin{figure}
	\centering
	\includegraphics[width=.87\textwidth]{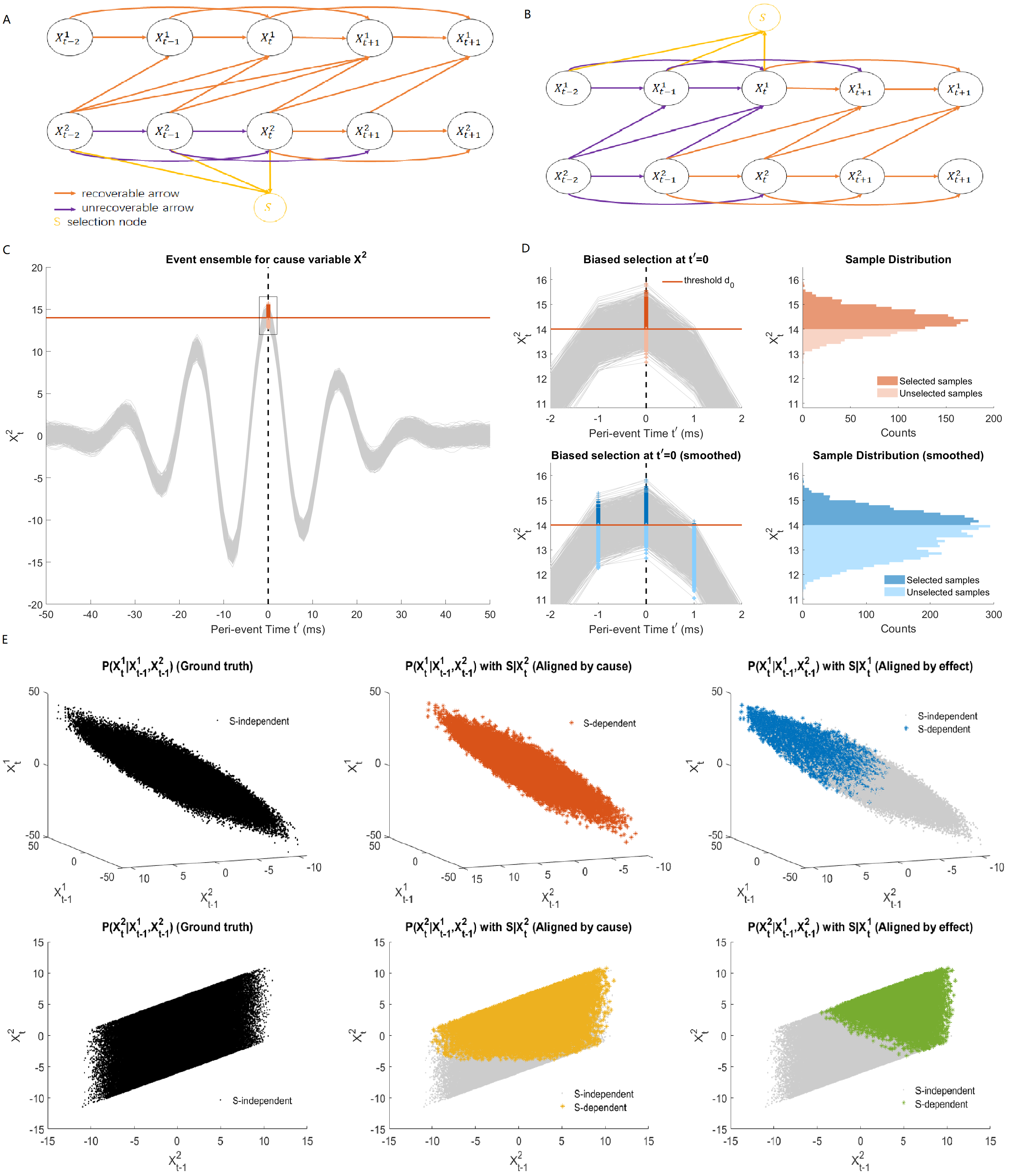}
	\caption{Illustration of selection bias due to thresholding and alignment. 
		(A) SCM of a bi-variate VAR(2) model with uni-directional coupling from $X^2$ to $X^1$ and a selection node S depending on states of the cause variable before peri-event time ($t^{\prime}<=0$). The selection node S represents partial selection of samples due to thresholding of the filtered cause signal (as the detection signal). Orange arrows makes the recoverable arrows with the current selection node, while purple arrows indicates the unrecoverable ones.  
		(B) The same SCM as in (A) with the selection node depending in a similar way on the effect signal.
		(C) An example event ensemble for the cause variable $X^2_t$ in (A-B) and the detection threshold.
		(D) Zoomed event ensenbles for (C) (left) and histograms for selected samples compared to the full sample (right). Upper panel illustrates selection bias at ground truth peri-event time $t^{\prime}=0$. 
		Lower panel shows selection bias at the peri-event time $t^{\prime}=0$ for detected events aligned by the peak. The aligned state $t^{\prime}=0$ is a smoothed state of neighboring states of the ground truth  $t^{\prime}=0$.
		(E) Illustration of recoverability when aligning by the cause. Subplots show joint distributions of the lagged variables and the putative effect variable of a VAR(1) model with uniformly distributed innovations, with left column for the ground-truth alignment, the middle column for aligning by the cause and right column for aligning by the effect. The conditional is only recoverable for the top middle panel. 
} \label{fig:bias}
\end{figure}

To model the effect of a threshold-based detection and alignment, based on \cite{bareinboim2012controlling} and \cite{bareinboim2014recovering}, we can modify the SCM in Figure~\ref{fig:scm}A to incorporate an additional node $S$ representing the selection variable, which is a binary variable indicating the time window is selected if and only if $S=1$ (see Supplementary Section A for background). Typically $S$ is defined by testing whether a continuous random variable goes over a predefined threshold. This continuous RV is itself a function of the time series values in a sliding time window, corresponding for example to a measure of the match between the time series and a predefined template. We can a priori choose $S$ to depend either on the cause variable $X^2_t$ (Figure~\ref{fig:bias}A) or on the effect variable $X^1_t$ (Figure~\ref{fig:bias}B). 
We look at the peri-event time $t^{\prime}$ relative to the reference time $t$. This selection based on variable $S$ is a priori not perfect, in the sense that it will not recover exactly the set of peri-event time series that we initially wish to analyze, i.e. those associated to a biologically relevant pattern of activity. Assuming that the detection method (e.g. the detection template) is well chosen, and the detection threshold is high enough, selection based on $S=1$ will typically ``over-select'', i.e. excluding some peri-event time series that would actually be relevant for our analysis. Figure~\ref{fig:bias}C(left, upper right) illustrates how thresholding selects only subset of peri-event trajectory samples at $t^{\prime}=0$ in a simulated scenario. 
This over-selection can then be modeled as sampling peri-event data from a conditional peri-event distribution $p(X|S)$, while we are interested in analyzing a ground truth distribution $p(X)$. This conditioning may induce a so-called \textit{selection bias} in the estimation of quantities we are interested in, notably the conditional distributions that enter the calculations of TE, DCS and rDCS. The impact of such bias on those quantities as been investigated in  \cite{bareinboim2012controlling,bareinboim2014recovering} within the SCM framework, as we describe in the following. 

For simplicity and consistency with the Results section, we will restrict ourselves to models with a unidirectional causal effect (either $X^1\to X^2$ or $X^2\to Z^1$) and that $S$ is only dependent on a finite number of negative peri-event times ($t^{\prime}\leq 0$) as in the case of a causal Finite Impulse Response (FIR) filter (for other cases, refer to Supplementary Section C.2. 
Figure~\ref{fig:bias}A, B illustrate in this setting that the causal arrow ($X^2\rightarrow X^1$) can be recovered at any peri-event time only when the selection node depends on the cause variable (see Supplementary Section C.2 for justification). Specifically, this means that 
$P\left(X_{t}^{1} \mid\boldsymbol{X}^1_{p,t}, \boldsymbol{X}^2_{p,t}, S\right) =P\left(X_{t}^{1} \mid \boldsymbol{X}^1_{p,t}, \boldsymbol{X}^2_{p,t}\right) $ 
for the SCM in Figure~\ref{fig:bias}A.  
For the opposite direction,  
$P\left(X_{t}^{2} \mid \boldsymbol{X}^1_{p,t},\boldsymbol{X}^2_{p,t}, S\right) \neq P\left(X_{t}^{2} \mid \boldsymbol{X}^2_{p,t}\right)$ for negative peri-event time $t^{\prime}\leq0$. 
For the case where $S$ depends on the effect variable, 
$P\left(X_{t}^{1} \mid\boldsymbol{X}^1_{p,t}, \boldsymbol{X}^2_{p,t}, S\right) \neq P\left(X_{t}^{1} \mid \boldsymbol{X}^1_{p,t}, \boldsymbol{X}^2_{p,t}\right) $ 
and 
$P\left(X_{t}^{2} \mid \boldsymbol{X}^1_{p,t},\boldsymbol{X}^2_{p,t}, S\right) \neq P\left(X_{t}^{2} \mid \boldsymbol{X}^2_{p,t}\right)$ 
for negative peri-event time $t^{\prime}\leq 0$ (see also Supplementary Section C.2. 
The $S$-dependent and $S$-independent conditionals are visualized in Figure~\ref{fig:bias}D for an example VAR(1) model, as described in Section~\ref{sec:te_insensit}, where the innovations $\eta^1_t$ and  $\eta^2_t$ are drawn from a uniform-distribution.
Similarly, the conditional model of the post-intervention scenario for rDCS with selection node depending on the cause
\[
\begin{aligned}	
	& p^{do(X^1_t \coloneqq f(\boldsymbol{X}^1_{p,t},\boldsymbol{X}^2_{p, t_\mathit{ref}}, \eta^1_t))}
	(X^1_t|\boldsymbol{X}^1_{p,t},\boldsymbol{X}^2_{p,t}, S)
	=\int P(\boldsymbol{X}^2_{p, t_\mathit{ref}}) \left(X_{t}^{1} \mid\boldsymbol{X}^1_{p,t}, \boldsymbol{X}^2_{p, t_\mathit{ref}}, S\right) \mathrm{d} \boldsymbol{X}^2_{p, t_\mathit{ref}}\\
	&=\int P(\boldsymbol{X}^2_{p, t_\mathit{ref}}) \left(X_{t}^{1} \mid\boldsymbol{X}^1_{p,t}, \boldsymbol{X}^2_{p, t_\mathit{ref}}\right) \mathrm{d} \boldsymbol{X}^2_{p, t_\mathit{ref}}
	=p^{do(X^1_t \coloneqq f(\boldsymbol{X}^1_{p,t},\boldsymbol{X}^2_{p, t_\mathit{ref}}, \eta^1_t))}
	(X^1_t|\boldsymbol{X}^1_{p,t},\boldsymbol{X}^2_{p,t})\\
\end{aligned}
\]
Therefore, the KL divergence in Eq.~\ref{eq:rDCS} can always be estimated correctly when selecting based on the cause values, while this does not hold for other directions or selecting on the effect. As rDCS is defined as the expectation over the KL divergence over the past states $\boldsymbol{X}^1_{p,t}$ and $\boldsymbol{X}^2_{p,t}$ (Eq.~\ref{eq:rDCS}), 
the estimated rDCS($X_t^2 \rightarrow X_t^1$) also depends on the unbiased sampling of the joint probability of $\boldsymbol{X}^1_{p,t}$ and $\boldsymbol{X}^2_{p,t}$. We argue here that this is normally satisfied as the threshold is set for a filtered detection signal but not for the original ones such that the latter are not affected.
In the most extreme case where the detection signal is the observation itself (as seen in Figure~\ref{fig:perturb_events}), the estimated rDCS($X_t^2 \rightarrow X_t^1$)  might be slightly amplified because the selected distribution tends to be biased towards larger values (also see Eq.~\ref{eq:rDCS}), while DCS($X_t^2 \rightarrow X_t^1$) might be underestimated according to Eq.~\ref{eq:DCS}. Thus its directionality is still preserved.

Next, we would like to clarify the relationship between the selection node, thresholding and alignment. Assume that there is a hidden state underlying the observed signals which are accessible from the data. The perfect alignment (considered \textit{ground truth}) refers to the condition where the hidden states are identical for all trials at each peri-event time $t^{\prime}$ in an extracted event ensemble, as shown in Figure~\ref{fig:bias}C(left) for $t^{\prime}=0$. 
In the $S$-dependent SCM, the aforementioned selection bias due to thresholding is confined for samples at each peri-event time $t^{\prime}$, as shown in Figure~\ref{fig:bias}C(upper right). We refer to this alignment situation as (events) aligned by \textit{single-time selection} of the selected variable.
However, as the hidden state is not known, by thresholding over the whole observed signal one often detects neighboring time points or points at neighboring local maxima (e.g., Figure~\ref{fig:bias}C(lower right)). 
Selecting all these points is smoothing or pooling over all these neighboring points, and the corresponding alignment is named as (events) aligned by \textit{smoothed selection} of the detected signal or variable. 
Notably, practically, a common way is to select the local peaks as the reference points, which can be understood as a non-uniform subsampling of the smoothed samples. Thus in the presented experiments, we will illustrate the results with the local peak alignment such that it is more accessible by the readers. 

Therefore, we propose that the selection bias-related recoverability theory can be applied to event ensembles aligned by local peaks over the threshold. 
In this case, while event ensembles are aligned by the true cause variable, the strength of connectivity (the arrow $X_t^2 \rightarrow X_t^1$) is not affected by thresholding (i.e., by the selection node) and so is the rDCS in the causal direction (rDCS($X_t^2 \rightarrow X_t^1$)). 
Although rDCS is biased for the other direction when aligned by the effect variable, for uni-directionally coupled systems the bias is small such that the contrast between two directions is preserved. 
As the true causal direction is unknown, we thus propose further that, to investigate the dominant causal direction between two event ensembles, we should focus on comparing the causality measures (TE, DCS and rDCS) for each direction when the events are aligned on the putative cause, i.e., $X_t^2 \rightarrow X_t^1|S(X^2)$ compared to $X_t^1 \rightarrow X_t^2|S(X^1)$.

\section{Results}

In this section, we first focus on illustrating the properties of TE, DCS and rDCS with simulated toy models. 
The problem of vanishing TE occurring with synchronized signals and the benefits of DCS in the same situation will be investigated in Section~\ref{sec:synchron_signals}. 
Next, we simulate a simple uni-directionally coupled VAR(4) system with rhythmic perturbations of the cause variable to generate transient events, where we will show that rDCS is able to reflect the change of causal effects due to this perturbation while TE and DCS fail. 
We also study the influence of the alignment method in the same example, as well as in empirical data of SWRs from uni-directionally coupled brain regions.

\subsection{The case of strongly-correlated signals} \label{sec:synchron_signals}

As mentioned in Section~\ref{sec:motivation}, TE does not capture well causal influences when the cause and effect signals are strongly correlated with each other, contray to DCS. 
Here, to illustrate such contrast, we simulate a bivariate dynamical system in the form of two 
synchronized continuous harmonic oscillators $x(t)$ and $y(t)$, with uni-directional coupling (i.e., $x(t)$ driving $y(t)$):
\begin{equation} \label{eq:driven_harmonic_oscillator}
\begin{cases}
\frac{d^2 x}{d t^2} &= -2\zeta_x \omega_x \frac{d x}{d t} - \omega^2_x x + n_x\,, \\
\frac{d^2 y}{d t^2} &= -2\zeta_y \omega_y \frac{d y}{d t} - \omega^2_y y + c x + n_y\,.
\end{cases}
\end{equation}

In this system, $x(t)$ is designed as an under-damped oscillator ($\zeta_x = 0.015722$) which approximately oscillates at  a period $T_x = 200$ samples corresponding to natural (angular) frequency $\omega_x = 2\pi/T_x = 0.0314$ rad/sample. 
To achieve synchrony, $y(t)$ is also designed as an under-damped oscillator ( $\zeta_y = 0.2$) whose intrinsic oscillation gradually vanishes and finally follows the oscillation of ${x(t)}$ with a coupling strength of $c = 0.098$. For $y(t)$, $T_y = 20$,  $\omega_y = 2\pi/T_y = 0.314$.
We also add small Gaussian innovations to both oscillators:  $n_x \sim \mathcal{N} (0, 0.02)$, $n_y \sim \mathcal{N} (0, 0.005)$. Adding this noise allows fitting a VAR model to the the signals to assess the causal interactions with TE and DCS.
VAR parameter estimation would fail with deterministic signals by causing the covariance matrix estimates to be singular. 

Using the Euler method with a time step of 1 and random initial points ($\mathcal{N}(0,1)$), we simulated 2000 trials of this uni-directionally coupled system with 1000-point length. We discarded the first 500 points to ensure that the time series reach a sufficient level of synchronization. We can see this system as a stationary VAR(2) process because numerical simulation with the Euler method generates data with its past two states. 

\begin{figure} 
	\includegraphics[width=\textwidth]{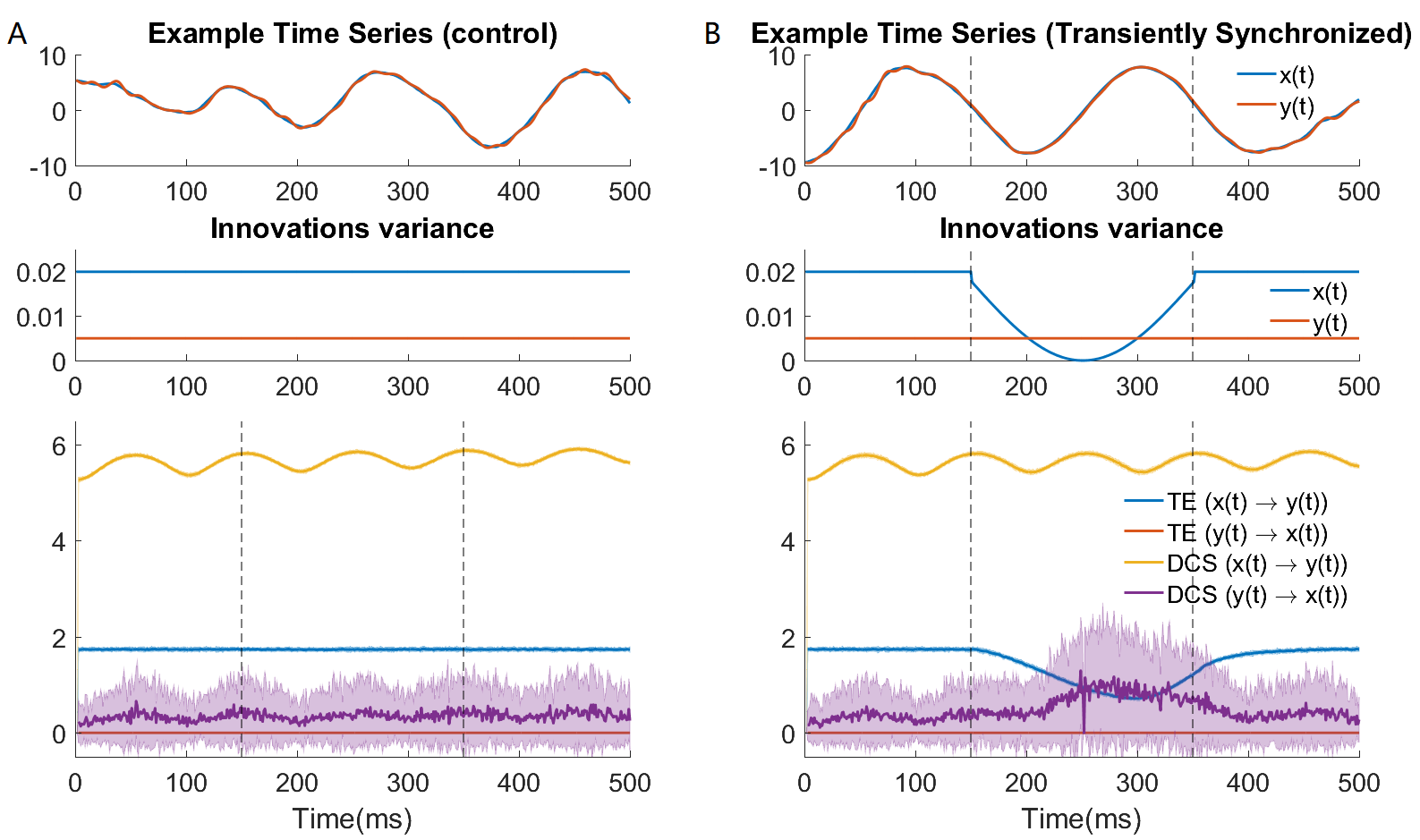}
	\caption{\label{fig:1} TE fails when the signals are strongly synchronized. 
		(A) Control experiments where synchrony is not changed.
		(top) Example trace of the bivariate signal in the control experiment. 
		(middle) Time-varying design of innovation's variance for both variables in the control experiment.
		(bottom) Time-varying TE and DCS results in the control experiment.
		(B) TE underperforms during transient increased synchrony induced by a tiny change in noise variance. The transient change can be seen as an event. Subfigure designs are the same as (A).}
\end{figure}

Figure~\ref{fig:1} (left panel) shows the results of time-varying TE and DCS for assessing the causal effects between $x(t)$ and $y(t)$. Calculation is performed in both the ground truth direction ($x(t) \rightarrow y(t)$) and the opposite direction. 
We first look at the control experiment. Consistent with the system's stationarity, TE is constant in both directions while being higher in the ground-truth direction. 
DCS in the ground-truth direction stays at a relatively high level, despite some small oscillation under a frequency similar to the intrinsic oscillation frequency of $x(t)$.

With respect to the detection of causal direction, both measures are able to detect the correct direction (i.e., causation for $x(t) \rightarrow y(t)$ is much larger than in the opposite direction). It is also reasonable that DCS in both directions is higher than TE, according to its definition in section~\ref{sec:DCS}. However, from the control experiment, we cannot conclude that the smaller TE values are due to its definition or due to the strong synchrony in the signals.

Therefore, we introduced a transient decrease of the noise variance in the cause signals ($x(t)$). The logic of designing this transient change is the following: the level of synchronization will increase with weaker noise, but the system and input magnitude remain the same because the contribution of the noise change to the signal amplitude is negligible; thus if TE is insensitive to the level of synchronization of signals, its values are expected to stay constant. However, as the results show in Figure~\ref{fig:1} (right panel), there is a transient decrease of TE during the interval where noise variance is decreased, suggesting that TE performs poorly in the cases where the cause and effect signals are strongly synchronized. 

\subsection{The case of deterministic perturbations} \label{sec:perturb_events}

In this section, we directly address the benefits of rDCS over TE and DCS when applied to signals driven by deterministic perturbations. 
To illustrate this specific property, we designed some simple transient events perturbing the innovation parameters of a stationary VAR process with uni-directional coupling. 
The events are generated by feeding the cause signal with innovations with non-zero time-varying means, such that both signals will exhibit temporal oscillations. 
We refer to these events as \textit{perturbation events} in the following sections. 
These perturbations intrinsically define a hidden state that parametrizes the ground truth distribution of peri-event data. We exploit the hidden state and demonstrate that the proposed alignment method in Section~\ref{sec:align} is efficient for recovering the time-varying causal direction between the two variables.

\subsubsection{Simulation procedure}\label{sec:perturb_simul}
\label{sec:TransEventsGen}

We simulated a non-stationary uni-directionally-coupled autoregressive system defined in Eq.~\ref{eq:full_model_x1_inhomo} and Supplementary Eq.S7. The causal direction is $X^2\rightarrow X^1$. The system is designed as a bivariate VAR($4$) process with a time-invariant coefficient matrix: $\mathbf{a}^\top=[-0.55, -0.45, -0.55, -0.85]$, $\mathbf{b}^\top=[1.4, -0.3, 1.5, 1.7]$, $\mathbf{c}^\top=[0, 0, 0, 0]$ and $\mathbf{d}^\top=[0.9, -0.25, 0, 0.25]$. These coefficients were randomly generated and and kept after checking the stability of the VAR($4$) system. Uni-directional interactions are ensured by setting the autoregressive coefficients associated to interactions in the opposite direction (i.e. $\mathbf{c}$) to zero for all lags. 

We enforce non-stationarity of $\eta^2_t$, the innovations of the ground truth cause process $\{X^2_t\}$. Both innovations $\eta^1_t$ and $\eta^2_t$ are drawn from a Gaussian distribution with unit variance (with no correlation in between, i.e., $\mathrm{Cov}[\eta^1_t,{\eta^2_t}^\top] = 0$); the difference is that $\mathbb{E}[\eta^1_t] = k^1_t = 0$ while $\mathbb{E}[\eta^2_t] = k^2_t$ is non-zero and time-varying. We designed the time-varying profile of $k^2_t$ as a Morlet-shaped waveform to mimic the oscillatory properties of neural event signals: $k^2_t = H \exp(-(\alpha x)^2/2) \cos(5 \alpha x)$, where $\alpha = 2/25$ is a constant controlling the event duration, and $H=4$ is the amplitude of the highest peak in the center of the event. The total duration of the Morlet-shaped waveform is 101 ms. The innovation's mean designed for $X^2$ is shown in Figure~\ref{fig:perturb_events}B (top left panel).

We generated this bi-variate VAR(4) process for 1300$s$ consisting of 5000 trials of perturbation events by transiently varying  $\eta^2_t$, detecting event occurrence based on the cause $X^2_t$, as illustrated in Figure~\ref{fig:perturb_events}A. 
The central peaks of these Morlet events are used as the ground-truth reference points for which peri-event time $t^{\prime}=0$, and used to extract a dataset of multi-trial events ensemble with a 200-$ms$ peri-event window such that $t^{\prime}$ range from -99$ms$ to +100$ms$  (i.e. there is no alignment procedure that could lead to selection bias, see Section~\ref{sec:align}). The event waveforms of the cause variable $X^2_t$ and the effect variable $X^1_t$ are illustrated in Figure~\ref{fig:perturb_events}B (bottom left, middle left).
The whole process is repeated 100 times to obtain variabilities plotted in the figure. 

\begin{figure} [b!]
	\centering
	\includegraphics[width=.8\textwidth]{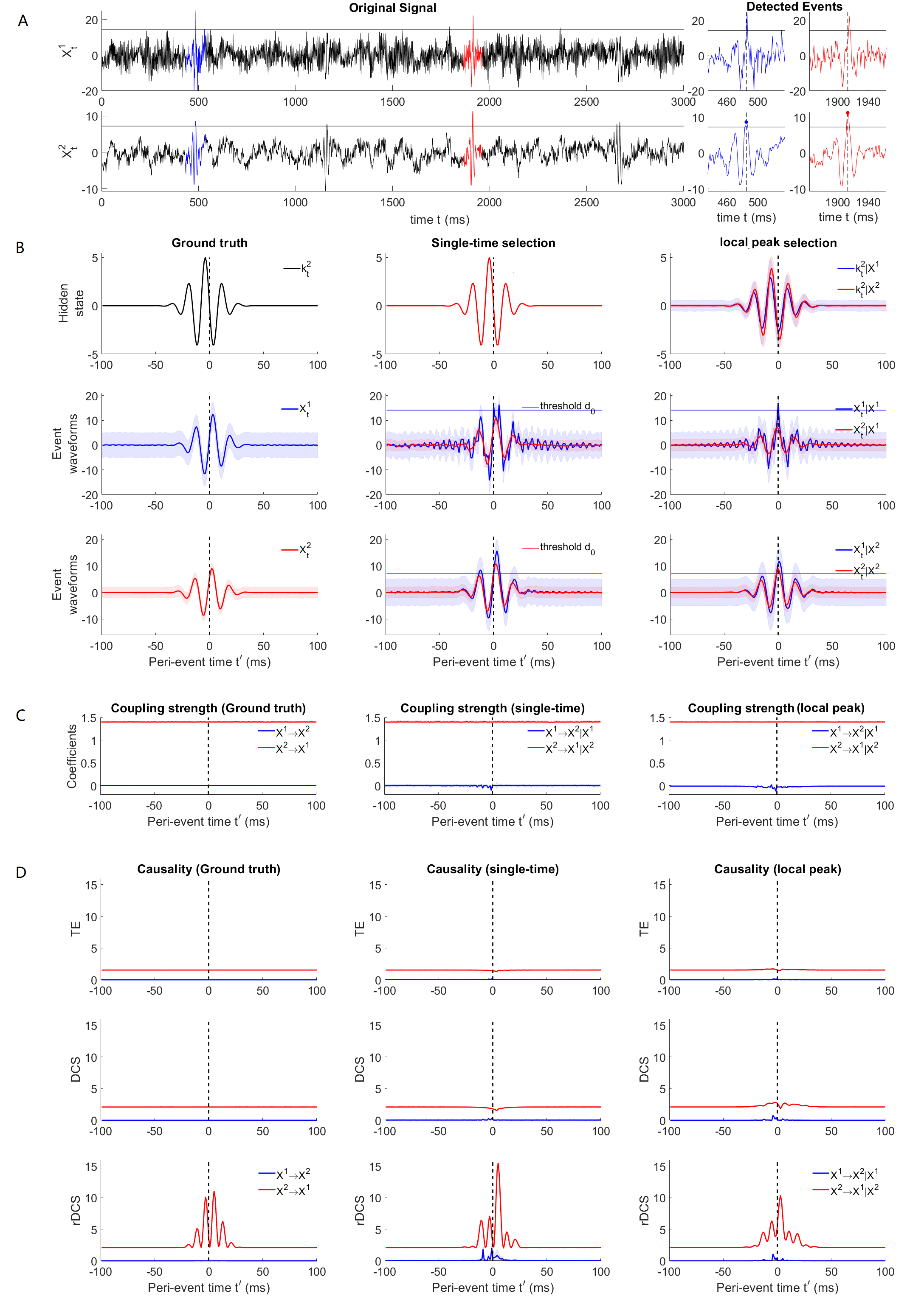}
	\caption{Causal analysis for simulated perturbation events with non-zero innovations. 
	(A) Example signal traces of the bi-variate VAR(4) system (black). Blue and red traces mark two example events detected by thresholding over the cause $X^2_t$. Blue and red dots show other reference points.
	(B) (Top) Hidden states for \textit{ground-truth alignment} (left), \textit{single time selection} of the ground truth event ensembles due to thresholding (middle) and 
	events aligned by local peaks over threshold (right). (Middle) ground
 } \label{fig:perturb_events}

\end{figure}
\addtocounter{figure}{-1}
\begin{figure} [t!]
	\caption{\textbf{(continued)} 
		truth event ensemble for $X_t^1$ (left) and bi-variate ensembles of the other two selections aligned by $X_t^1$ (middle, right). Thin blue line represents the threshold in $X_t^1$.
		(Bottom) Same settings as in (middle) but aligned by $X^2_t$.
		(C) Example elements of coupling strength in the ground truth directions $X^2_t \rightarrow X^1_t$ (red) and the opposite direction $X^1_t \rightarrow X^2_t$ (blue) for 3 types of event ensembles aligned by putative cause.
		(D) TE (left), DCS (middle) and rDCS (right) for all 3 types of event ensembles aligned by putative cause.}
\end{figure}

\subsubsection{Effect of trial alignment on model estimation and causality measures}

As the innovation is designed as deterministic (i.e., identical) across trials, with this alignment, the event ensembles obtained by the ground truth reference points define a dataset with \textit{ground-truth alignment}. We can compare the  VAR model estimation and causality measures resulting from this dataset to the outcomes obtained by aligning events based on the detection of either variable $X^2$ or $X^1$, as discussed in Section~\ref{sec:align}. 

To validate the theory related to the recoverability of ground truth conditional probabilities in the presence of selection bias due to the event detection procedure, 
we test the \textit{single-time selection} case where biased selection is only performed for samples at peri-event time $t^{\prime}=0$ (as Figure~\ref{fig:bias}C(upper right)), thus not changing the hidden state alignment (Figure~\ref{fig:perturb_events}B(upper middle)). 
The reference points are decided as all ground truth reference points higher than a threshold $d_0$, where $d_0$ is 3 times the standard deviation of the whole signal. 
The selected event ensembles of both variables are shown in Figure~\ref{fig:perturb_events}A(middle middle) for thresholding over $X_t^1$ and Figure~\ref{fig:perturb_events}A(middle right) for thresholding over $X_t^2$.
Notably, this kind of selection is only feasible when the hidden state in known, which is not realistic practically for real data.

Therefore, to demonstrate the appropriateness of the proposed alignment (i.e., aligning by all data points of one variable over the threshold, or the local peaks
), we assume the signal itself as the detection signal and $d_0$ as the threshold. We obtain an event ensemble by selecting
local peaks 
for points over 3 times $d_0$ as new reference points, which is shown in Figure~\ref{fig:perturb_events}B(middle right and bottom right).
This can be seen as a smoothed version of the real events, which is also confirmed by checking the aligned hidden states (Figure~\ref{fig:perturb_events}B(upper right)).

While inferring VAR model parameters of  the event ensembles according to \cite{KaidiBIC}, the true model order (4) can be recovered for all five ensembles. 
Figure~\ref{fig:perturb_events}C demonstrates the recoverability of conditional models for ensembles aligned by the putative cause.
One of the 4 coupling strengths from the putative cause to the putative effect is plotted as red curves.
As described in the simulation procedure in Section~\ref{sec:perturb_simul}, the coupling strength is constant over time, which is reflected in Figure~\ref{fig:perturb_events}C(left). 
Consistent with the theory in Section~\ref{sec:align}, biased selection of event trials on the samples at $t^{\prime}=0$ leads to unbiased estimation of the coupling strength $X_t^2 \rightarrow X_t^1$ aligned by the cause $X^2$ (denoted also as ``$\mid X^2$'' in Figure~\ref{fig:perturb_events}C(middle)). 
By comparison, the coupling strength in the other direction is slightly biased at negative peri-event times ($t^{\prime}$) but still relatively close to its true value (0). 
This contrast holds for alignment with local peaks 
over threshold, as seen in Figure~\ref{fig:perturb_events}C(right). 

Figure~\ref{fig:perturb_events}D(upper, middle, lower) shows the corresponding results of how causality measures perform in the three alignment scenarios. 
During the periods where no transient events occur, all three measures are able to infer a time-invariant causal effect in the ground-truth direction ($X^2\rightarrow X^1$) compared to the opposite direction. 
Besides, in line with theoretical predictions, DCS is higher than TE and is equal to rDCS. 
During the perturbation events, in the ground truth direction TE and DCS remain constant and rDCS exhibit a rhythmic pattern. These results match the theoretical predictions: TE and DCS measures the connectivity strength, which does not change, while rDCS measures the combined causal effect related to the connectivity and the event-based changes at the cause while yielding larger variations transmitted to the effect node. 

In the case where event ensembles are aligned by \textit{single-time selection} of the cause $X^2_t$, TE and DCS of the ground truth direction is underestimated while rDCS is slightly overestimated around $t^{\prime}=0$, which is consistent with the theories in Section~\ref{sec:align}. A bias appears in the opposite direction while aligned by the effect, but the direction is detected correctly.
The case of local peak alignment shows similar results, except the peak amplitude of the smoothed rDCS is less amplified. 

Thus, this simulational experiment of perturbation events demonstrates the effectiveness of rDCS in reflecting the causal influence when the cause is perturbed by a deterministic exogenous input compared to TE and DCS, validating that rDCS is a better measure to address event-based causal interactions. 
We show that in practice, detecting via thresholding and aligning the event ensemble with 
the local peaks of the putative cause  
is a good way to recover the ground truth event-based causality given uni-directional connections: the trick is to address the coupling when the events are aligned by the putative cause, then a specific feature of invariance in the presence of selection bias can be applied.
This property will be further tested on real data in the next section.

\subsection{Validation on SWRs-based causality between CA3 and CA1 regions}\label{sec:rodent_ripples}

Sharp Wave-Ripple (SWR) events, hypothesized as a key element in implementing memory consolidation in the brain, have been reported in the electrophysiological recordings within the hippocampus of both macaques and rodents. 
In this section we detect SWRs in an experimental dataset to investigate the behavior of TE, DCS and rDCS in a neuroscientific context where the event-hosting brain regions are uni-directionally coupled. 

SWRs are primarily generated in the CA1 area of the hippocampus. The somas of CA1 pyramidal cells are located in the in the pyramidal layer ('pl') while their dendritic trees are rooted in the stratum radiatum ('sr'). It is hypothesized that the dendritic trees receive strong excitatory inputs from the pyramidal cells in CA3 which generate post-synaptic activities in the dendritic trees. This results in LFP activities in low frequencies (0-30Hz, due to the sharp-wave) and in gamma band (30-80Hz, due to CA3 oscillations). Then the dendritic activities propagate to the soma, where recurrent interactions between inhibitory and excitatory cells generate a very fast oscillation, the ripples (80-250Hz). 

\begin{figure}
	
	\includegraphics[width=\textwidth]{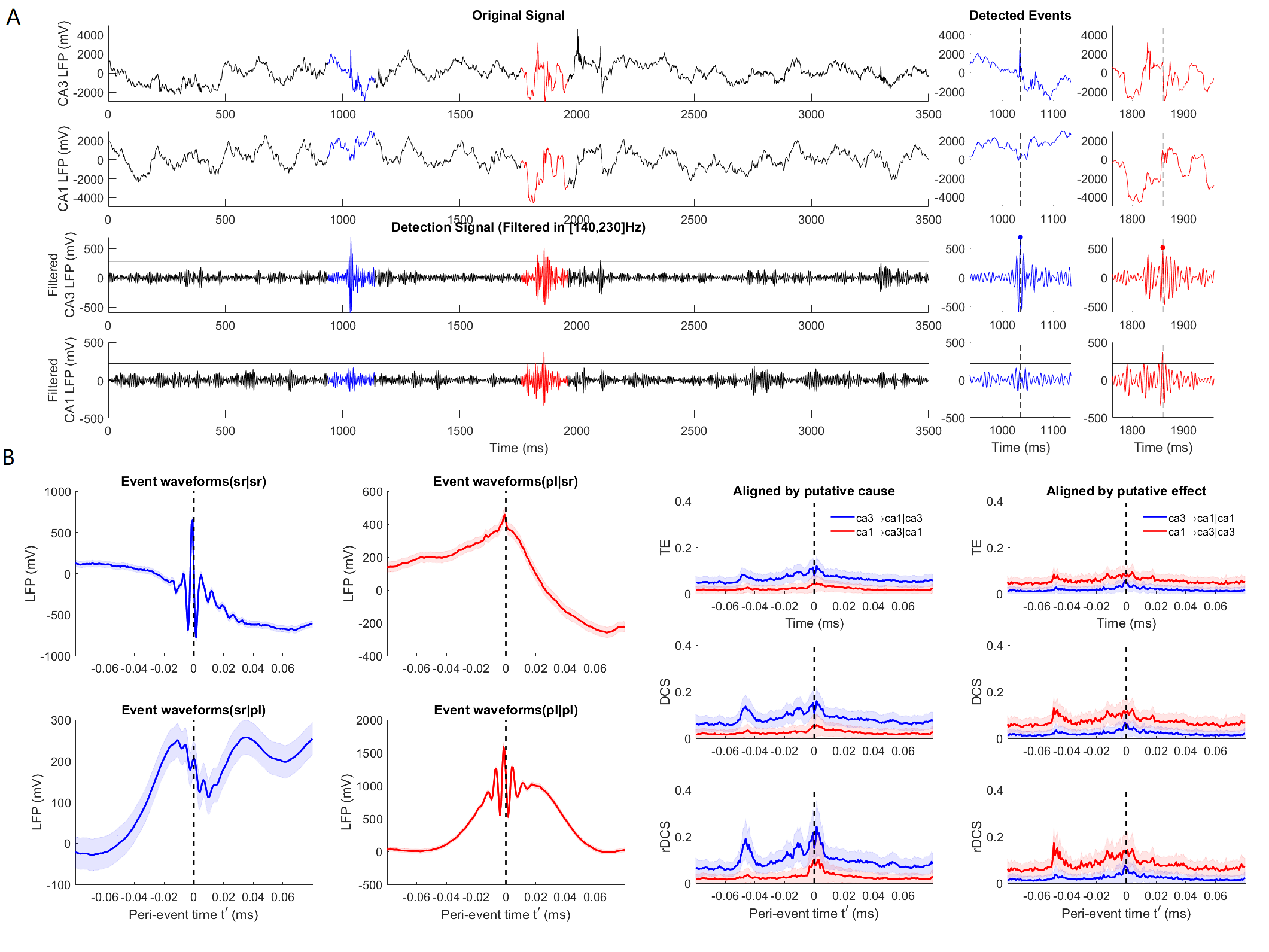}
	
	\caption{ Event-based causal analysis for SWRs in rodent hippocampal CA3 and CA1 regions.
		(A) Examples signal traces of the original signals and bandpass filtered signals of CA3 and CA1 regions (black). Blue and red traces mark two example events detected by thresholding over the cause CA3 and aligned by the local peak. Blue and red dots show other reference points.
		(B) Event waveforms of SWR event ensembles at CA3 (left) and CA1 (right) regions aligned by CA3 (upper) and CA1 (lower) signals.
		Shades repensent the ensemble standard error averaged over 1024 channel pairs.
		(C) Peri-event causality measured by TE (upper), DCS (middle) and rDCS (lower) for event ensembles aligned by the putative cause (left) and the putative effect (right). Shades reflect standard deviation of 100 repeated bootstrapped ensembles.
	}\label{fig:ripple_ca3ca1}
	
\end{figure}

We applied the event-based causality analysis to an open source dataset where electrophysiological recordings in the CA3 and CA1 regions of rodent hippocampus have been performed with 4 shanks of 8 channels simultaneously in each region \citep{mizuseki2014neurosharing}. 
In agreement with the SWR generation mechanism explained in the above paragraph, anatomical studies \citep{CSICSVARI2000585} 
support uni-directional anatomical coupling between these two regions within the hippocampal formation, i.e., CA3$\rightarrow$CA1. 
The analysis is based on two Local Field Potential (LFP) data sessions recorded from the rat named 'vvp01' with a sampling rate of 1252Hz. 
An example trace of a channel pair of both CA3 and CA1 regions is shown in Figure~\ref{fig:ripple_ca3ca1}A. 
{As SWRs are more challenging to observe during behavioral sessions, 
we perform our analysis only on a session of sleep which lasts 4943.588$s$.} 

Following \cite{mizuseki2009theta}, we detect SWRs by applying an 49-ordered FIR filter in the frequency band [140,230]Hz to each channel of signals in both regions. 
Similar to Section~\ref{sec:perturb_events}, we set a threshold over the mean of the filtered signals (5 SD) to locate the events and align them according to the local peak points over threshold in the filtered signals of either region in a channel pair. 

Figure~\ref{fig:ripple_ca3ca1}A(lower) shows the case aligned on the CA3 signals. 
The peri-event window for display has been chosen to be [-119.8, 119.8]$ms$, while VAR model estimation and the BIC-based model order selection are performed according to \citep{KaidiBIC}.
For each channel pair, we obtain two bi-variate event ensembles, thus extracting 2*1024 ensembles for all channel pairs for each alignment condition. 
The event waveforms and statistics of an example channel pair for different alignments are illustrated in Figure~\ref{fig:ripple_ca3ca1}B.

SWR-based causality measures shown in Figure~\ref{fig:ripple_ca3ca1}C compare the alignment by the putative cause and by the putative effect.
{The reference states used for estimating rDCS are the averaged states over the first 16$ms$ time points in the window.}
The standard deviation plotted in the figure originates from 100 times bootstrapped ensembles and the variability is averaged over 1024 channel pairs. 
In line with the theoretical predictions, the ground truth direction (CA3$\rightarrow$CA1) is well recovered when using an alignment by the putative cause, but not when aligning by the putative effect. 
{TE, DCS and rDCS in the opposite of the truth direction are not significantly different from zero, which is consistent with the uni-directionality of anatomical connections posited by anatomical studies. } 
{Significantly stronger causal influences in the ground truth direction are shown by TE, DCS and rDCS before the alignment point ($t^{\prime}=0$), matching the hypothesized SWR generating mechanism that the CA3 region drives the SWR interactions in CA1 region.} 
{The lack of difference between the two directions in more stationary states might be explained by the ineffectivity of causal measures based on linear VAR models to capture non-linearity \citep{shajarisales2015telling}}.
{The transient increase in the non-ground truth direction when using alignment on the putative cause might be explained by the selection bias elaborated on in Section~\ref{sec:align}.}

\section{Discussion}

In summary, we have discussed the benefits and shortcomings of two time-varying causality measures (TE and DCS) in characterizing causal interactions based on peri-event data.
To address their insensitivity to deterministic perturbations, we proposed a novel measure, the rDCS, that implements an intervention on both the cause and the mechanism in the SCM framework.
We compared the performance of these causality measures on perturbation events with innovations with time-varying means and
and electrophysiological recordings of hippocampal SWRs. 
The benefits of rDCS is supported by the perturbation events presented in Section~\ref{sec:perturb_events}.
As causality analysis of transient events aims at uncovering the network mechanisms underlying these phenomena (e.g., addressing whether one event \textit{drives} the other), we argue for the use of rDCS as it provably captures causal influences due to event-related changes in the cause that propagate to target regions through anatomical connections, even if these changes have little variability across trials.  
	
Our results of the performance of rDCS are tested with SWRs events recorded within the hippocampus. However, potentially this measure can be applied to all multiple trial event pairs recorded in two brain regions, for example, ponto-geniculate-occipital waves simultaneously recorded in the thalamus and cortex. The effect of time-varying non-zero-meaned innovations might reflect the exogenous variables un-included in the model, and thus might be helpful in understanding the mechanism of underlying event generation.

However, we show that the data preparation procedure (i.e., the detection procedure and the alignment of the events) potentially affects the detection of causal effects and the quantification of their strength, consistent with predictions of \citep{bareinboim2012controlling} and \citep{bareinboim2014recovering}. 
In particular, while the directionality of peri-event causal interactions is expected to be ab intrinsic property of the underlying mechanisms, we showed that the inferred causal direction is dependent on the alignment methods (i.e., the event-triggering region). 
We hypothesize that this is due to selection bias, which is supported by our theoretical investigations. 

Actually, aligning neural events based on the activity in a single region is a common practice in event-related brain research. Our results thus suggest that causality analysis of such peri-event data collection methods may be affected by a selection bias, which should be controlled for. 
We have demonstrated empirically that for uni-directional coupling, computing rDCS on peri-event datasets triggered by the putative cause is normally effective in revealing the true causal direction.  
Thus, as a future direction, it is imperative to devise a correction approach for the selection bias such that peri-event dynamics can be accurately recovered to achieve a better characterization of causal influences between transient neural events.

\section*{Conflict of Interest Statement}
The authors declare that the research was conducted in the absence of any commercial or financial relationships that could be construed as a potential conflict of interest.

\section*{Author Contributions}
Conceptualization, M.B. and N.K.; Methodology, M.B. and K.S.; Software, K.S.; Validation, K.S.; Formal Analysis, K.S. and M.B.;  Resources, N.K.;  Writing—Original Draft Preparation, K.S. and M.B.; Writing—Review \& Editing, K.S., and M.B.; Supervision, M.B.

\section*{Funding}
This work is based on the projects supported by the International Center for Primate Brain Research (Departmental Funding E114N61A21). This work was supported by the German Federal Ministry of Education and Research (BMBF): Tübingen AI Center, FKZ: 01IS18039B. 

\section*{Acknowledgments}
M.B. and K.S. would like to thank P. Geiger and M. Yang for useful discussions.

\section*{Data Availability Statement}
The datasets generated and analyzed for this study can be found in the GitHub project 
\url{https://github.com/KaidiShao/event_causality_frontiers}. The rat hippocampal Sharp Wave-Ripples datasets during sleep are found on \url{https://crcns.org/data-sets/hc/hc-3/about-hc-3}  (session vvp01-4-9, date 2006-4-9\_18-43-47).

\bibliographystyle{unsrtnat}
\bibliography{exported-references-multiTrialGranger}

\end{document}